\documentclass[aps,pre,showpacs,notitlepage,amsmath,amssymb]{revtex4-1}
\usepackage{graphicx}
\usepackage{wasysym}

\begin{document}

\title{Influence of external potentials on heterogeneous diffusion processes}

\author{Rytis Kazakevi\v{c}ius}
\email{rytis.kazakevicius@tfai.vu.lt}
\affiliation{Institute of Theoretical Physics and Astronomy, Vilnius University,
Saul\.{e}tekio 3, LT-10222 Vilnius, Lithuania }

\author{Julius Ruseckas}
\affiliation{Institute of Theoretical Physics and Astronomy, Vilnius University,
Saul\.{e}tekio 3, LT-10222 Vilnius, Lithuania }

\begin{abstract}
In this paper we consider heterogeneous diffusion processes with the
power-law dependence of the diffusion coefficient on the position
and investigate the influence of external forces on the resulting
anomalous diffusion. The heterogeneous diffusion processes can yield
subdiffusion as well as superdiffusion, depending on the behavior
of the diffusion coefficient. We assume that not only the diffusion
coefficient but also the external force has a power-law dependence
on the position. We obtain analytic expressions for the transition
probability in two cases: when the power-law exponent in the external
force is equal to $2\eta-1$, where $2\eta$ is the power-law exponent
in the dependence of the diffusion coefficient on the position, and
when the external force has a linear dependence on the position. We
found that the power-law exponent in the dependence of the mean square
displacement on time does not depend on the external force, this force
changes only the anomalous diffusion coefficient. In addition, the
external force having the power-law exponent different from $2\eta-1$
limits the time interval where the anomalous diffusion occurs. We
expect that the results obtained in this paper may be relevant for
a more complete understanding of anomalous diffusion processes.
\end{abstract}


\maketitle

\section{Introduction}

There are many systems and processes where the time dependence of
the centered second moment is not linear as in the classical Brownian
motion. Such family of processes is called anomalous diffusion. In
one dimension the anomalous diffusion is characterized by the power-law
time dependence of the mean square displacement (MSD) \cite{Bouchaud1990}
\begin{equation}
\langle(\Delta x)^{2}\rangle=K_{\alpha}t^{\alpha}\,.\label{eq:anomalous-diff}
\end{equation}
Here $K_{\alpha}$ is the anomalous diffusion coefficient. When $\alpha\neq1$,
this time dependence deviates from the linear function of time characteristic
for the Brownian motion. If $\alpha<1$, the phenomenon is called
subdiffusion. Occurrence of subdiffusion has been experimentally observed,
for example, in the behavior of individual colloidal particles in
two-dimensional random potential energy landscapes \cite{Evers2013}.
It has been theoretically shown that active particles moving at constant
speed in a heterogeneous two-dimensional space experience self trapping
leading to subdiffusion \cite{Chepizhko2013}. Usually it is assumed
that subdiffusive behavior is caused by the particle being trapped
in local minima for prolonged times before it escapes to a neighboring
minima. Continuous time random walks (CTRWs) with on-site waiting-time
distributions falling slowly as $t^{-\alpha-1}$ predict a subdiffusive
behavior \cite{Metzler2000,Schubert2013}. 

Superdiffusion processes, characterized by the nonlinear dependence
(\ref{eq:anomalous-diff}) of the MSD on time with the power-law exponent
in the range $1<\alpha<2$, constitute another subclass of anomalous
diffusion processes. Superdiffusion is observed, for example, in vibrated
dense granular media \cite{Scalliet2015}. Theoretical models suggest
that supperdiffusion can be caused by L\'evy flights \cite{Metzler2000}.
L\'evy flights resulting in a superdiffusion can be modeled by fractional
Fokker-Planck equations \cite{Fogedby1994} (or Langevin equations
with an additive L\'evy stable noise). In many experimental studies
it is only possible to show that signal intensity distribution has
L\'evy law-tails: distribution function of turbulent magnetized plasma
emitters \cite{Marandet2003} and step-size distribution of photons
in hot vapors of atoms \cite{Mercadier2009} have L\'evy tails. This
indirectly shows that in these systems superdiffusion could be found.

Anomalous diffusion does not uniquely indicate the processes occurring
in the system, because there are different stochastic processes sharing
the behavior of the MSD (\ref{eq:anomalous-diff}). The physical mechanisms
leading to the deviations from the linear time dependence of the MSD
can depend on the system or on the temporal and spatial ranges under
consideration. For example, diffusion described by CTRW has been observed
for sub-micron tracers in biological cells \cite{Tabei2013,Weigel2011,Jeon2011},
structured coloidal systems \cite{Wong2004} and for charge carrier
motion in amorphous semiconductors \cite{Scher1975,Schubert2013}.
Fractional Brownian motion and fractional Langevin equations has been
used to model the dynamics in membranes \cite{Jeon2012,Kneller2011},
motion of polymers in cells \cite{Kepten2013}, tracer motion in complex
liquids \cite{Szymanski2009,Jeon2013}. Diffusion of even smaller
tracers in biological cells has been described by a spatially varying
diffusion coefficient \cite{Kuehn2011}.

Recently, in Refs.~\cite{Cherstvy2013,Cherstvy2013a,Cherstvy2014a,Cherstvy2014}
it was suggested that the anomalous diffusion can be a result of heterogeneous
diffusion process (HDP), where the diffusion coefficient depends on
the position. Spatially dependent diffusion can occur in heterogeneous
systems. For example, heterogeneous medium with steep gradients of
the diffusivity can be created in thermophoresis experiments using
a local variation of the temperature \cite{Maeda2012,Mast2013}. Mesoscopic
description of transport in heterogeneous porous media in terms of
space dependent diffusion coefficients is used in hydrology \cite{Haggerty1995,Dentz2010}.
In turbulent media the Richardson diffusion has been described by
heterogeneous diffusion processes \cite{Richardson1926}. Power-law
dependence of the diffusion coefficient on the position has been proposed
to model diffusion of a particle on random fractals \cite{OShaughnessy1985,Loverdo2009}.
In bacterial and eukaryotic cells the local cytoplasmic diffusivity
has been demonstrated to be heterogeneous \cite{English2011,Kuehn2011}.
Motion of a Brownian particle in an environment with a position dependent
temperature has been investigated in Ref.~\cite{Kazakevicius2015}.
In random walk description the spatially varying diffusion coefficient
can be included via position dependence of the waiting time for a
jump event \cite{Srokowski2014}, the position dependence occurs because
in the heterogeneous medium the properties of a trap can reflect the
medium structure. This is the case for diffusion on fractals and multifractals
\cite{Schertzer2001}. Inhomogeneous versions of continuous time random
walk models for water permeation in porous ground layers were proposed
in Ref.~\cite{Dentz2012}. Heterogeneous diffusion process might
be applicable to describe anomalous diffusion in such systems.

The goal of this paper is to consider HDPs with the power-law dependence of the
diffusion coefficient on the position and to analytically investigate the
influence of external forces on the resulting anomalous diffusion. The influence
of the external forces on HDPs has not been systematically analyzed. In this
paper we assume that the forces are characterized by a power-law dependence on
the position. Such forces can arise in various systems: In many cases the
potentials causing the deterministic forces are power-law functions of position,
for example linear and harmonic potentials. Power-law potential with arbitrary
power-law exponent acting on a nanoparticle have been created in
Ref.~\cite{Cohen2005}. Logarithmic potentials yielding forces behaving as
$x^{-1}$ have been applied to describe dynamics of particles near a long,
charged polymer \cite{Manning1969}, momentum diffusion in dissipative optical
lattices \cite{Sagi2012}, long-range interacting systems \cite{Bouchet2005},
bubbles in DNA molecules \cite{Wu2009}. As we demonstrate, the external force
having a certain value of the power-law exponent does not restrict the region of
diffusion and change only the anomalous diffusion coefficient without changing
the scaling exponent $\alpha$. Other values of the power-law exponent in the
external force can lead to the exponential cut-off of the probability density
function (PDF) and restrict the region of diffusion, limiting the time interval
when the anomalous diffusion occurs. We expect that the results obtained in this
paper may be relevant for a more complete understanding of anomalous diffusion
processes.

The paper is organized as follows: In Sec.~\ref{sec:hdp} we summarize
the main properties of HDPs with the power-law dependence of the diffusion
coefficient on the position. The influence of an external force with
a particular value of the power-law exponent is investigated in Sec.~\ref{sec:drift}.
This value of the power-law exponent is the same as in the drift correction
for transformation from the Stratonovich to the It\'o stochastic
equation. We consider external forces having other values of the power-law
exponent in Sec.~\ref{sec:boundaries}. Sec.~\ref{sec:concl} summarizes
our findings.

\section{Free heterogeneous diffusion process\label{sec:hdp}}

Heterogeneous diffusion process with the power-law dependence of the
diffusion coefficient on the position, introduced in Ref.~\cite{Cherstvy2013},
is described by the Langevin equation
\begin{equation}
dx=\sigma|x|^{\eta}\circ dW_{t}\,.\label{eq:hdp}
\end{equation}
Here $\eta$ is the power-law exponent of multiplicative noise, $\sigma$
is the amplitude of noise and $W_{t}$ is a standard Wiener process
(Brownian motion). This stochastic differential equation (SDE) is
interpreted in Stratonovich sense. For mathematical convenience and
for further generalization we transform Eq.~(\ref{eq:hdp}) to the
It\^o convention:
\begin{equation}
dx=\frac{1}{2}\sigma^{2}\eta|x|^{2(\eta-1)}xdt+\sigma|x|^{\eta}dW_{t}\,.
\label{eq:hdp-ito}
\end{equation}
The Fokker-Planck equation corresponding to the SDE (\ref{eq:hdp})
is \cite{Gardiner2004}
\begin{equation}
\frac{\partial}{\partial t}P(x,t)=\frac{1}{2}\sigma^{2}\frac{\partial}{\partial x}
\left[|x|^{\eta}\frac{\partial}{\partial x}\left(|x|^{\eta}P(x,t)\right)\right]\,.
\label{eq:FP-hdp}
\end{equation}
With the reflective boundaries at small positive $x=x_{\mathrm{min}}$
and large $x=x_{\mathrm{max}}$, Eq.~(\ref{eq:FP-hdp}) leads to
the steady-state PDF $P_{0}(x)\sim x^{-\eta}$. Without such boundaries
the time-dependent solution of Eq.~(\ref{eq:FP-hdp}) with the initial
condition $P(x,0)=\delta(x)$ is given by a stretched (when $\eta>0$)
or compressed (when $\eta<0$) Gaussian \cite{Cherstvy2013}
\begin{equation}
P(x,t)=\frac{|x|^{-\eta}}{\sqrt{2\pi\sigma^{2}t}}
\exp\left(-\frac{|x|^{2(1-\eta)}}{2(1-\eta)^{2}
\sigma^{2}t}\right)\,.\label{eq:sol-1}
\end{equation}
When $\eta<1$, this solution describes exponential cut-off at large
values of $x$ and power-law behavior at small values of $x$. The
position of the cut-off moves towards large values of $x$ with increase
of time $t$. In contrast, when $\eta>1$ the solution (\ref{eq:sol-1})
describes exponential cut-off at small values of $x$ and power-law
behavior at large values of $x$. The position of cut-off moves towards
smaller values of $x$ with increase of time $t$.

In Ref.~\cite{Cherstvy2013} it has been demonstrated that Eq.~(\ref{eq:hdp})
(or, equivalently, Eq.~(\ref{eq:hdp-ito})) leads to the power-law
time dependence of the MSD 
\begin{equation}
\langle x^{2}(t)\rangle\sim(\sigma^{2}t)^{\frac{1}{1-\eta}}\,.\label{eq:scaling-1}
\end{equation}
This behavior of the MSD means that HDP described by Eq.~(\ref{eq:hdp})
yields superdiffusion for $1>\eta>0$ and subdiffusion for $\eta<0$.
For $\eta>0$ the diffusivity increases with increasing $x$, leading
to a progressive acceleration of the diffusing particle. In contrast,
the diffusivity decreases with increasing $x$ when $\eta<0$. For
$\eta>1$ the particle becomes localized. When $\eta=1$, the MSD
grows not as a power-law of time, but exponentially \cite{Lau2007,Fulinski2011}.

The HDP~(\ref{eq:hdp}) displays weak non-ergodicity, that is the
scaling of time and ensemble averages is different. Specifically,
in Ref.~\cite{Cherstvy2013} it has been shown that the average over
the trajectories
\begin{equation}
\left\langle \overline{\delta^{2}(\Delta)}\right\rangle =
\frac{1}{N}\sum_{i=1}^{N}\overline{\delta_{i}^{2}(\Delta)}
\end{equation}
of the the time-averaged MSD
\begin{equation}
\overline{\delta^{2}(\Delta)}=\frac{1}{T-\Delta}\int_{0}^{T-\Delta}[x(t+\Delta)-x(t)]^{2}dt
\end{equation}
scales as
\begin{equation}
\left\langle \overline{\delta^{2}(\Delta)}\right\rangle
\sim\frac{\Delta}{T^{\frac{\eta}{\eta-1}}}\,.
\end{equation}
Thus time-averaged MSD depends on the time difference $\Delta$ linearly,
in contrast to the power-law behavior of MSD in Eq.~(\ref{eq:scaling-1}).

Another interesting property of HDPs is the behavior of the distribution
of the time-averaged MSD $\overline{\delta^{2}}$ of individual realizations.
When $\eta<0$, the distribution of $\overline{\delta^{2}}$ decays
to zero at $\overline{\delta^{2}}=0$ \cite{Cherstvy2014}. This behavior
of the distribution in HDPs is different than the behavior in CTRWs,
where there is a finite fraction of immobile particles resulting in
the finite value of the distribution at $\overline{\delta^{2}}=0$.
This difference allows to distinguish between different origins of
anomalous diffusion.

\section{External force that does not limit the anomalous diffusion\label{sec:drift}}

In general, not only a random force leading to the diffusion but also
a deterministic drift force can be present in the Langevin equation.
Therefore, the question arises how external force influences the anomalous
diffusion described by heterogeneous diffusion process. To investigate
this question we will consider Eq.~(\ref{eq:hdp-ito}) with an additional
drift term. As the results obtained in this Section show, for the
appearance of the anomalous diffusion it is sufficient to consider
only positive values of $x$. Therefore from now on we will write
just $x$ instead of the absolute value $|x|$, assuming the presence
of the boundary that does not allow for the diffusing particle to
enter the region $x\leqslant0$.

Due to the power-law dependence of the diffusion coefficient on the position,
the drift term also being a power-law function of the position allows us to
obtain analytical expressions. The external forces characterized by power-law
dependence on the position can be created experimentally, for example, a
power-law potential with arbitrary power-law exponent acting on a nanoparticle
have been created in Ref.~\cite{Cohen2005}. First of all, in this Section we
will consider the case where the external force has the same dependence on
coordinate $x$ as the drift correction for the transformation from the
Stratonovich to the It\'o stochastic equation. Such a drift term can arise not
only due to an external force but can also represent a noise-induced drift
\cite{Volpe2016}. Thus we will generalize the SDE~(\ref{eq:hdp-ito}) to take the
form
\begin{equation}
dx=\sigma^{2}\left(\eta-\frac{\nu}{2}\right)x^{2\eta-1}dt
+\sigma x^{\eta}dW_{t}\,.\label{eq:sde-1}
\end{equation}
Here $\nu$ is a new parameter describing the additional drift term.
The meaning of the parameter $\nu$ is as follows: when the reflective
boundaries at small positive $x=x_{\mathrm{min}}$ and large $x=x_{\mathrm{max}}$
are present, the steady-state PDF is a power-law function of position
with the power-law exponent $\nu$, $P_{0}(x)\sim x^{-\nu}$. Comparison
of Eq.~(\ref{eq:sde-1}) with Eq.~(\ref{eq:hdp-ito}) shows that
the free HDP is obtained when $\nu=\eta$.

The SDE~(\ref{eq:sde-1}) has been proposed in
Refs.~\cite{Kaulakys2004,Kaulakys2006} to generate signals having $1/f^{\beta}$
noise in a wide range of frequencies. Such nonlinear SDEs have been applied to
describe signals in socio-economical systems \cite{Gontis2010,Mathiesen2013} and
as a model of neuronal firing \cite{Ton2015}. According to
Ref.~\cite{Kaulakys2006}, the power-law exponent $\beta$ in the power spectral
density $S(f)\sim f^{-\beta}$ is related to the parameters of SDE
(\ref{eq:sde-1}) as 
\begin{equation}
\beta=1+\frac{\nu-3}{2(\eta-1)}\,.
\end{equation}

For some values of the parameter $\nu$ the SDE~(\ref{eq:sde-1}) can be obtained
from Eq.~(\ref{eq:hdp}) by changing the prescription for calculating stochastic
integrals. Let us consider the SDE
\begin{equation}
dx=\sigma x^{\eta}\circ_{\gamma}dW_{t}\,,\label{eq:hdp-gamma}
\end{equation}
together with the interpretation of stochastic integrals as
\cite{Karatzas2012,Volpe2016}
\begin{equation}
\int_{0}^{T}f(x(t))\circ_{\gamma}dW_{t}
=\lim_{N\rightarrow\infty}\sum_{n=0}^{N-1}f(x(t_{n}))\Delta W_{t_{n}}\,,\qquad
t_{n}=\frac{n+\gamma}{N}T\,.
\end{equation}
The parameter $\gamma$, $0\leqslant\gamma\leqslant 1$, defines the prescription
for calculating stochastic integrals. Commonly used values of the parameter
$\gamma$ are $\gamma=0$ corresponding to pre-point It\^o convention,
$\gamma=1/2$ corresponding to mid-point Stratonovich convention and $\gamma=1$
corresponding to post-point H\"anggi-Klimontovich \cite{Hanggi1982,Klimontovich1994}, 
kinetic or isothermal convention \cite{Pesce2013,dosSantos2010,Sokolov2010}. The integration
convention should be determined from the experimental data or derived from
another model \cite{vanKampen1981}. The SDE (\ref{eq:hdp-gamma}) also has been investigated in Ref.~\cite{Heidernatsch2015}. Transformation of Eq.~(\ref{eq:hdp-gamma}) to
the It\^o equation yields \cite{Karatzas2012,Kwok2012,Arenas2014}
\begin{equation}
 dx=\sigma^{2}\gamma(\eta-1)x^{2\eta-1}dt+\sigma x^{\eta}dW_{t}\,.
\end{equation}
This equation has the form of SDE~(\ref{eq:sde-1}) with the parameter $\nu$
being
\begin{equation}
\nu=2[\gamma+(1-\gamma)\eta]\,.
\end{equation}
Since $0\leqslant\gamma\leqslant 1$, the range of possible values of the
parameter $\nu$ obtained by changing the prescription in Eq.~(\ref{eq:hdp}) is
limited: $2\eta\leqslant\nu\leqslant2$ when $\eta<1$ and
$2\leqslant\nu\leqslant2\eta$ when $\eta>1$. In this Section we do not place
such restriction on the possible values of $\nu$. The values of $\nu$ outside of
this range can be obtained due to the action of an external force.

The Fokker-Planck equation corresponding to the SDE~(\ref{eq:sde-1}) is
\cite{Gardiner2004}
\begin{equation}
\frac{\partial}{\partial t}P(x,t)=
\sigma^2\left(\frac{\nu}{2}-\eta\right)
\frac{\partial}{\partial x}[x^{2\eta-1}P(x,t)]
+\frac{\sigma^2}{2}\frac{\partial^2}{\partial x^2}[x^{2\eta}P(x,t)]\,.
\end{equation}
The time-dependent PDF of the process given by Eq.~(\ref{eq:sde-1})
can be obtained as follows: Transformation of the variable $x$ to
a new variable $y=x^{1-\eta}$ (assuming that $\eta\neq1$) leads
to the SDE
\begin{equation}
dy=-\frac{1}{2}\sigma^{\prime2}\nu^{\prime}\frac{1}{y}dt
+\sigma^{\prime}dW_{t}\,,\label{eq:bessel}
\end{equation}
where
\begin{equation}
\nu^{\prime}=\frac{\eta-\nu}{\eta-1}\,,\qquad\sigma^{\prime}=|\eta-1|\sigma\,.
\end{equation}
Equation (\ref{eq:bessel}) has the form of a Bessel process
\cite{Jeanblanc2009}. This connection with the Bessel process has also been
pointed out in Ref.~\cite{Heidernatsch2015}. The known analytic form of the
solution of the Fokker-Planck equation
\begin{equation}
\frac{\partial}{\partial t}P_{y}=\frac{1}{2}\sigma^{\prime2}\nu^{\prime}
\frac{\partial}{\partial y}y^{-1}P_{y}+\frac{1}{2}\sigma^{\prime2}
\frac{\partial^{2}}{\partial y^{2}}P_{y}\label{eq:FP-Bessel}
\end{equation}
corresponding to SDE (\ref{eq:bessel}) is \cite{Jeanblanc2009,Bray2000,Martin2011}
\begin{equation}
P(y,t|y_{0},0)=\frac{y^{\frac{1-\nu^{\prime}}{2}}
y_{0}^{\frac{1+\nu^{\prime}}{2}}}{\sigma^{\prime2}t}
\exp\left(-\frac{y^{2}+y_{0}^{2}}{2\sigma^{\prime2}t}\right)
I_{-\frac{\nu^{\prime}+1}{2}}
\left(\frac{yy_{0}}{\sigma^{\prime2}t}\right)\,.
\label{eq:sol-bessel}
\end{equation}
Here $I_{n}(z)$ is the modified Bessel function of the first kind. This PDF
obeys the initial condition $P(y,t=0|y_{0},0)=\delta(y-y_{0})$. For
completeness, one possible way of obtaining this solution of
Eq.~(\ref{eq:FP-Bessel}) is presented in Appendix~\ref{sec:bessel}. The PDF
(\ref{eq:sol-bessel}) can be normalized and represents an acceptable solution
only when $-\frac{\nu^{\prime}+1}{2}>-1$, that is when $\nu^{\prime}<1$. When
this inequality is not satisfied, the Bessel process leads to a total absorption
at the origin in a finite time \cite{Karlin1981}. Transforming the variables
back we obtain the time-dependent PDF satisfying the initial condition
$P(x,t=0|x_{0},0)=\delta(x-x_{0})$: 
\begin{equation}
P(x,t|x_{0},0)=\frac{x^{\frac{1-2\eta-\nu}{2}}
x_{0}^{\frac{1-2\eta+\nu}{2}}}{|\eta-1|\sigma^{2}t}
\exp\left(-\frac{x^{2(1-\eta)}+x_{0}^{2(1-\eta)}}{2(\eta-1)^{2}\sigma^{2}t}\right)
I_{\frac{\nu+1-2\eta}{2(\eta-1)}}\left(\frac{x^{(1-\eta)}
x_{0}^{(1-\eta)}}{(\eta-1)^{2}\sigma^{2}t}\right)\,.
\label{eq:pdf-x}
\end{equation}
Similarly as Eq.~(\ref{eq:sol-bessel}), the solution (\ref{eq:pdf-x}) is valid
when $\frac{\nu+1-2\eta}{2(\eta-1)}>-1$. This condition is equivalent to $\nu>1$
and $\eta>1$ or $\nu<1$ and $\eta<1$. Similar expression of the PDF
(\ref{eq:pdf-x}) has been obtained in Ref.~\cite{Heidernatsch2015} by
considering the HDP with various values of the prescription parameter $\gamma$.

Using Eq.~(\ref{eq:pdf-x}) we can calculate the time-dependent average
of a power of $x$:
\begin{eqnarray}
\langle x^{a}\rangle_{x_{0}} & = & \int_{0}^{\infty}x^{a}P(x,t|x_{0},0)dx\nonumber \\
 & = & \frac{\Gamma\left(\frac{\nu-1-a}{2(\eta-1)}\right)}{\Gamma\left(\frac{\nu-1}{2(\eta-1)}\right)}
\left(2(\eta-1)^{2}\sigma^{2}t\right)^{\frac{a}{2(1-\eta)}}\,{}_{1}F_{1}\left(\frac{a}{2(\eta-1)};
\frac{\nu-1}{2(\eta-1)};-\frac{x_{0}^{2(1-\eta)}}{2(\eta-1)^{2}\sigma^{2}t}\right)\,.
\label{eq:avg-x-a}
\end{eqnarray}
Here $_{1}F_{1}(a;b;z)$ is the Kummer confluent hypergeometric function. The
integral is finite when: i) $\eta>1$ and $\nu>1+a$ with $a>0$ or $\nu>1$ with
$a<0$; ii) $\eta<1$ and $\nu<1+a$ with $a<0$ or $\nu<1$ with $a>0$.
Eq.~(\ref{eq:avg-x-a}) for the $a$-th moment of $x$ has been derived also in
Ref.~\cite{Heidernatsch2015}. In particular, the average of $x$ is equal to
\begin{equation}
\langle x\rangle_{x_{0}}=
\frac{\Gamma\left(\frac{\nu-2}{2(\eta-1)}\right)}{\Gamma\left(\frac{\nu-1}{2(\eta-1)}\right)}
\left(2(\eta-1)^{2}\sigma^{2}t\right)^{\frac{1}{2(1-\eta)}}\,{}_{1}F_{1}\left(\frac{1}{2(\eta-1)};
\frac{\nu-1}{2(\eta-1)};-\frac{x_{0}^{2(1-\eta)}}{2(\eta-1)^{2}\sigma^{2}t}\right)
\label{eq:avg-x}
\end{equation}
and is finite when $\nu>2$ and $\eta>1$ or $\nu<1$ and $\eta<1$.
The average of the square of $x$ is equal to
\begin{equation}
\langle x^{2}\rangle_{x_{0}}=
\frac{\Gamma\left(\frac{\nu-3}{2(\eta-1)}\right)}{\Gamma\left(\frac{\nu-1}{2(\eta-1)}\right)}
\left(2(\eta-1)^{2}\sigma^{2}t\right)^{\frac{1}{1-\eta}}\,{}_{1}F_{1}\left(\frac{1}{(\eta-1)};
\frac{\nu-1}{2(\eta-1)};-\frac{x_{0}^{2(1-\eta)}}{2(\eta-1)^{2}\sigma^{2}t}\right)
\label{eq:avg-x-2}
\end{equation}
and is finite when $\nu>3$ and $\eta>1$ or $\nu<1$ and $\eta<1$.
When time $t$ is large, that is when 
\begin{equation}
\frac{x_{0}^{2(1-\eta)}}{2(\eta-1)^{2}\sigma^{2}t}\ll1\,,
\label{eq:cond-t-large}
\end{equation}
the hypergeometric function in Eq.~(\ref{eq:avg-x-a}) is approximately
equal to $1$. Thus for large time the average $\langle x^{a}\rangle_{x_{0}}$
does not depend on the initial position $x_{0}$ and is a power-law
function of time: 
\begin{equation}
\langle x^{a}\rangle_{x_{0}}\approx
\frac{\Gamma\left(\frac{\nu-1-a}{2(\eta-1)}\right)}{\Gamma\left(\frac{\nu-1}{2(\eta-1)}\right)}
\left(2(\eta-1)^{2}\sigma^{2}t\right)^{\frac{a}{2(1-\eta)}}\,.
\end{equation}
As a consequence we obtain that for large time $t$ satisfying the condition
(\ref{eq:cond-t-large}) the average of the square of $x$ depends on time as
$t^{1/(1-\eta)}$. In addition, using Eqs.~(\ref{eq:avg-x}) and
(\ref{eq:avg-x-2}) we get that the the variance $\langle(x-\langle
x\rangle)^{2}\rangle=\langle x^{2}\rangle-\langle x\rangle^{2}$ has the same
dependence on time: $\langle(x-\langle x\rangle)^{2}\rangle\sim t^{1/(1-\eta)}$.
We can conclude, that not only the original HDP equation (\ref{eq:hdp}), but
also the equation (\ref{eq:sde-1}) with additional power-law force exhibit
anomalous diffusion. The power-law exponent in the time dependence is the same
as in Eq.~(\ref{eq:scaling-1}), the external force considered in this Section
does not change the character of the anomalous diffusion as long as the
condition (\ref{eq:cond-t-large}) is satisfied. The power-law exponent
$1/(1-\eta)$ depends only on the random force in the SDE (\ref{eq:sde-1}) and
does not depend on the parameter $\nu$ characterizing the deterministic external
force. The external force changes only the anomalous diffusion coefficient.

\begin{figure}
\includegraphics[width=0.33\textwidth]{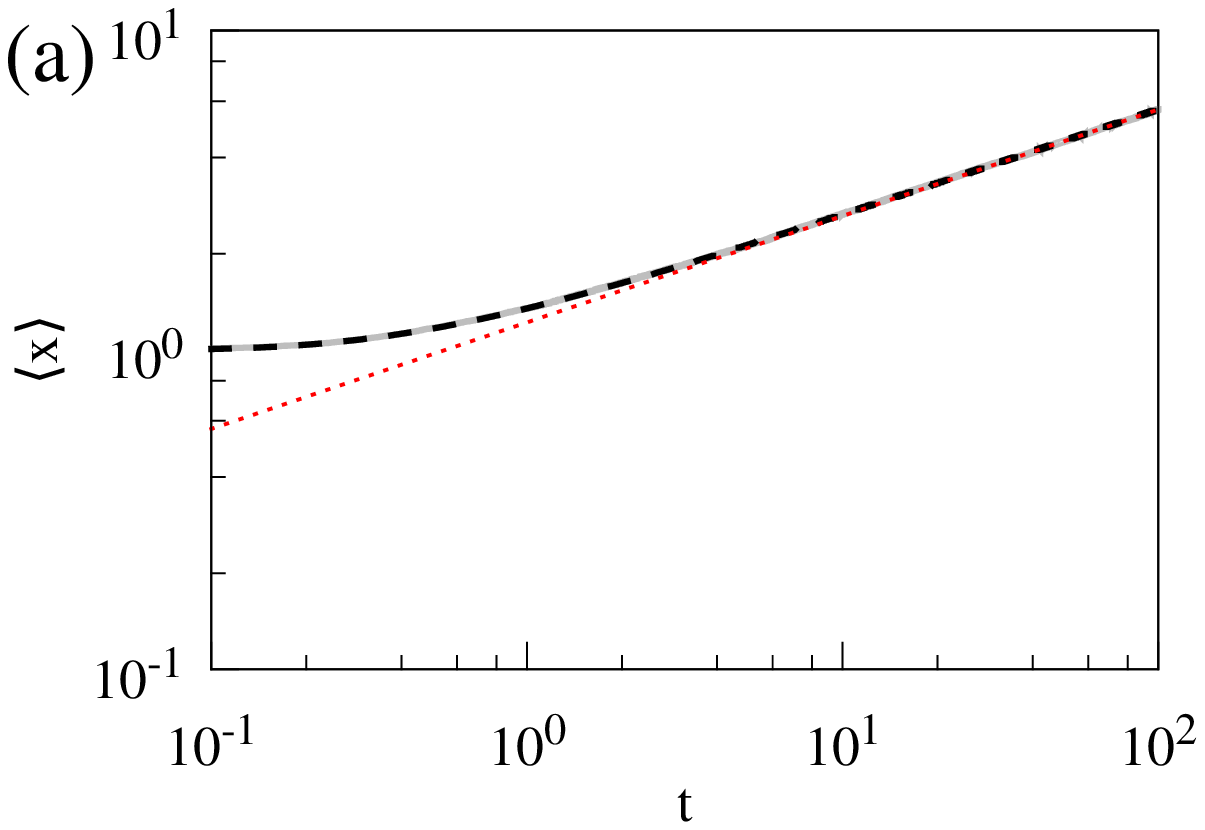}\includegraphics[width=0.33\textwidth]{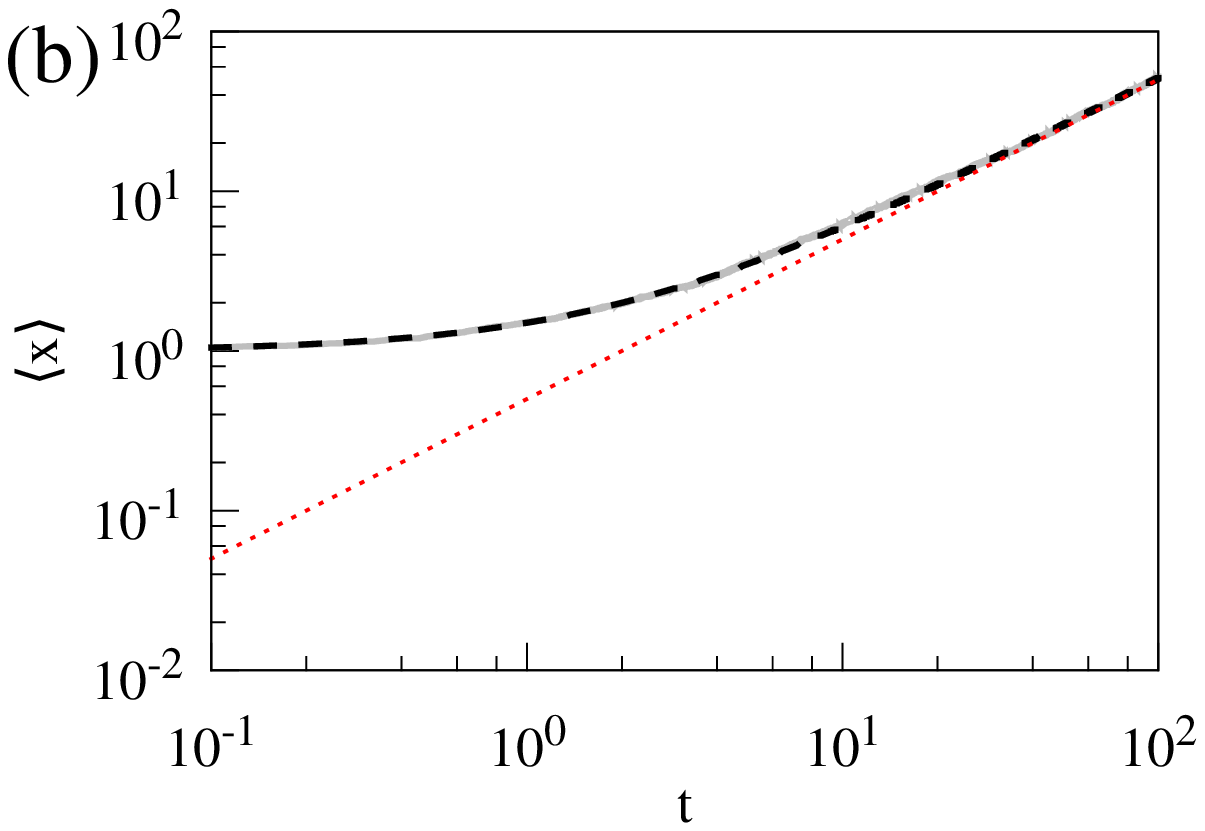}\includegraphics[width=0.33\textwidth]{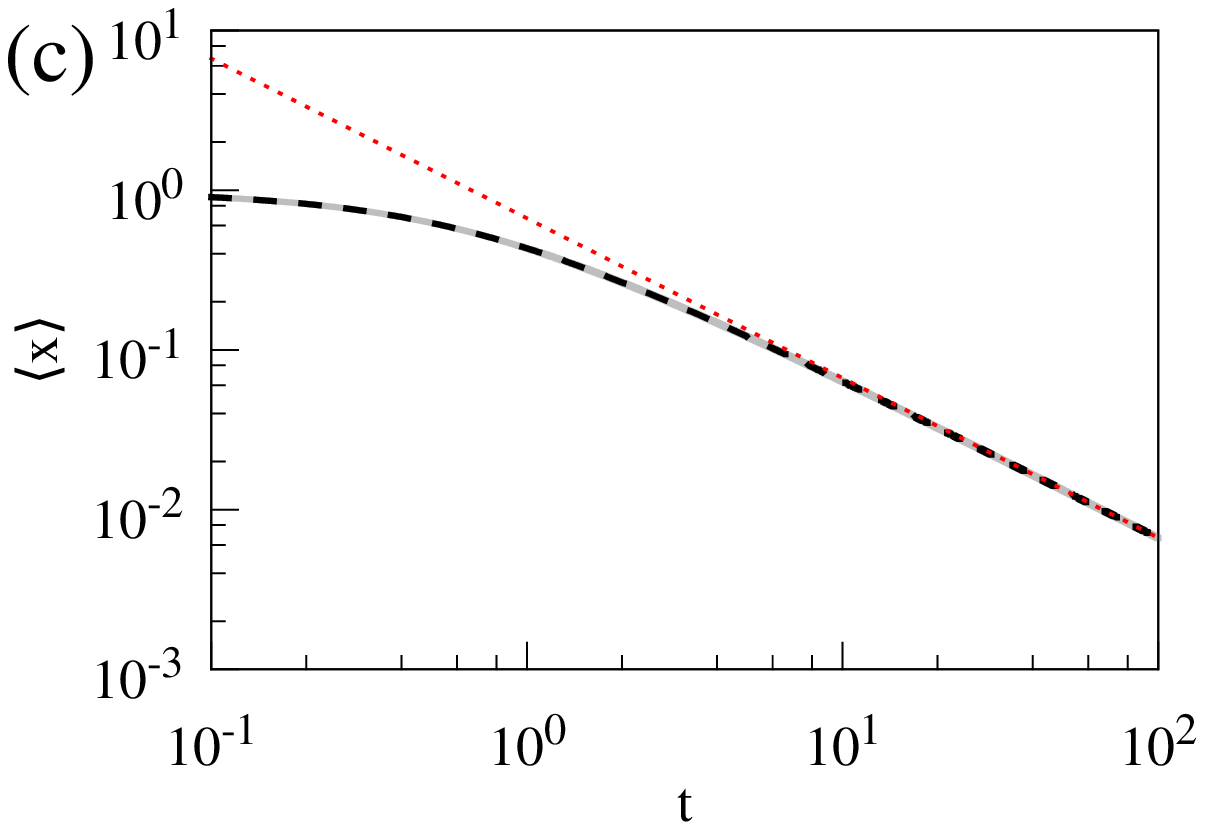}\\
\includegraphics[width=0.33\textwidth]{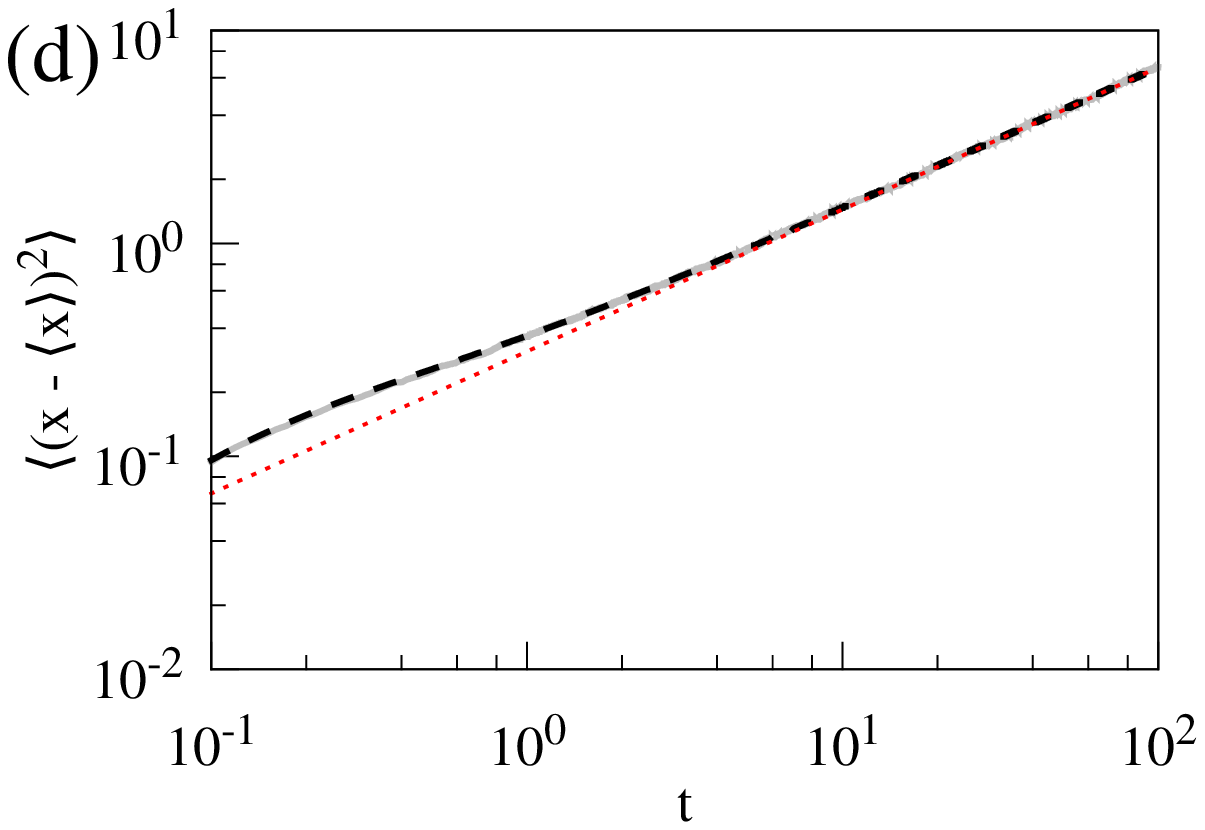}\includegraphics[width=0.33\textwidth]{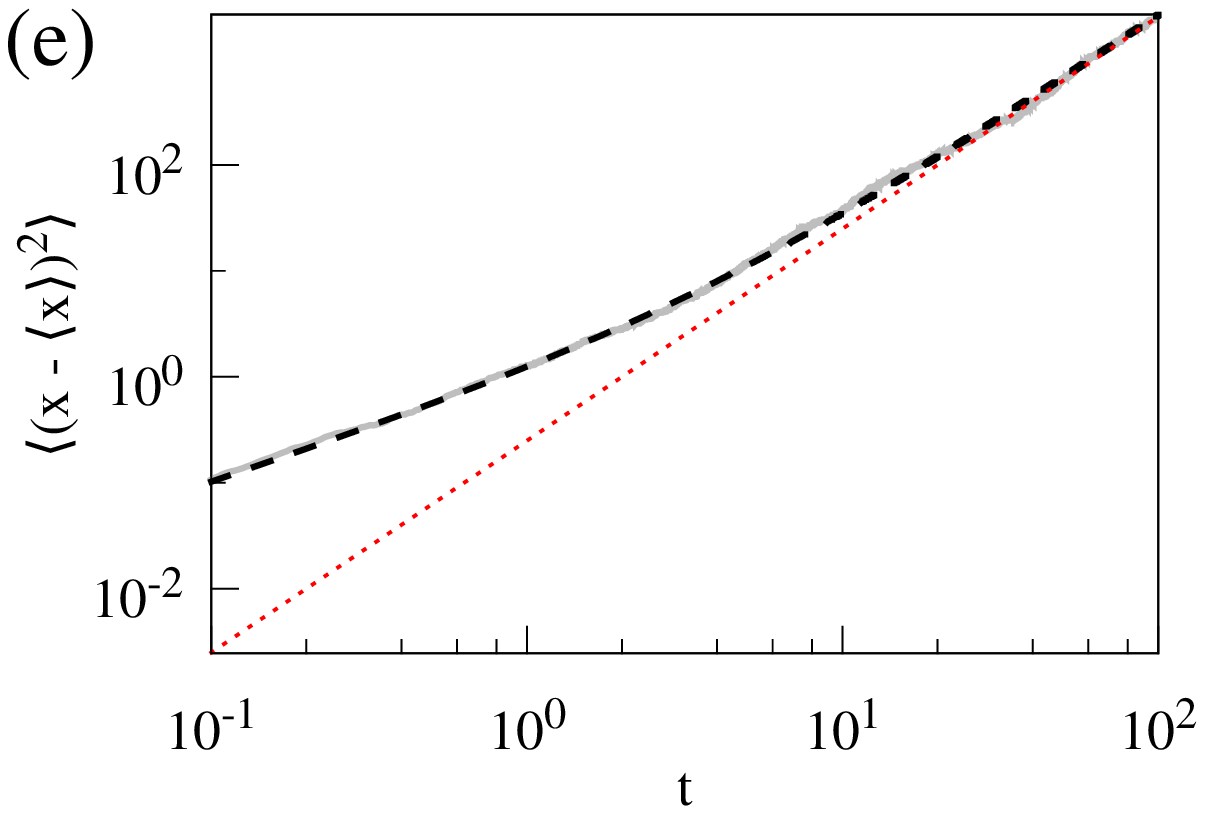}\includegraphics[width=0.33\textwidth]{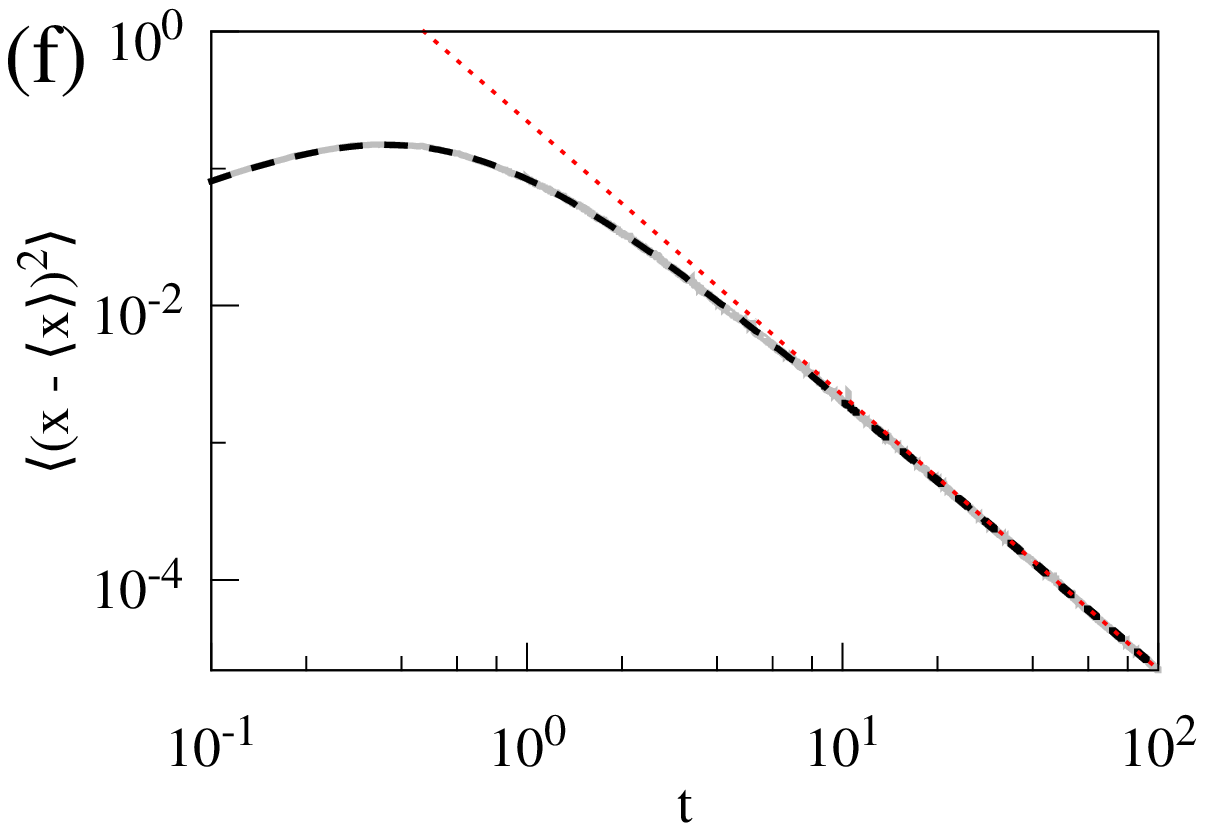}
\caption{Dependence of the mean (a,b,c) and variance (d,e,f) on time for various
values of the parameters $\eta$ and $\nu$ when the position of the diffusing
particle changes according to Eq.~(\ref{eq:sde-1}). Solid gray lines show
numerical result, dashed black lines are calculated using Eqs.~(\ref{eq:avg-x})
and (\ref{eq:avg-x-2}), dotted (red) lines show the power-law dependence on time
$\sim t^{1/[2(1-\eta)]}$ for (a,b,c) and $\sim t^{1/(1-\eta)}$ for (d,e,f). The
parameters are $\sigma=1$ and $\eta=-\frac{1}{2}$, $\nu=-1$ for (a,d);
$\eta=\frac{1}{2}$, $\nu=0$ for (b,c); $\eta=\frac{3}{2}$, $\nu=5$ for (c,f).
The initial position is $x_{0}=1$.}
\label{fig:mean-var}
\end{figure}

We check the analytic results obtained in this Section by comparing them with
numerical simulations. The power-law form of the coefficients in SDE
(\ref{eq:sde-1}) allows us to introduce an operational time $\tau$ in addition
to the physical time $t$ so that the diffusion coefficient in the operational
time becomes constant \cite{Ruseckas2015}. The relation between the physical
time $t$ and the operational time $\tau$ is specified by the equation
$dt=\sigma^{-2}x^{-2\eta}d\tau$. For the numerical solution of
SDE~(\ref{eq:sde-1}) we use the Euler-Maruyama scheme with a variable time step
$\Delta t_{k}=\Delta\tau/(\sigma^{2}x_{k}^{2\eta})$ which is equivalent to the
introduction of the operational time \cite{Ruseckas2015}. Thus the numerical
method of solution of SDE~(\ref{eq:sde-1}) is given by the equations
\begin{eqnarray}
x_{k+1} & = & x_{k}+\left(\eta-\frac{\nu}{2}\right)\frac{\Delta\tau}{x_{k}}
+\sqrt{\Delta\tau}\varepsilon_{k}\,,\\
t_{k+1} & = & t_{k}+\frac{\Delta\tau}{\sigma^{2}x_{k}^{2\eta}}\,.
\end{eqnarray}
Here $\Delta\tau\ll1$ is the time step in the operational time and
$\varepsilon_{k}$ are normally distributed uncorrelated random variables with a
zero expectation and unit variance. To avoid the divergence of the diffusion and
drift coefficients at $x=0$ in the numerical simulation, we insert a reflective
boundary at $x=10^{-3}$ . This modification is analogous to the regularization
of the diffusivity at $x=0$, performed in
Refs.~\cite{Cherstvy2013,Cherstvy2014}. When $\eta>1$ and $\nu>1$ the PDF of $x$
increases for small values of $x$. In this case the operational time introduced
in such a way that the coefficient before noise becomes proportional to the
first power of $x$ can lead to a more efficient numerical method. Such an
operational time is introduced by the equation
$dt=\sigma^{-2}x^{-2(\eta-1)}d\tau$ and the numerical method of solution becomes
\begin{eqnarray}
x_{k+1} & = & x_{k}\left[1+\Delta\tau\left(\eta-\frac{\nu}{2}\right)
  +\sqrt{\Delta\tau}\varepsilon_{k}\right]\,,\\
t_{k+1} & = & t_{k}+\frac{\Delta\tau}{\sigma^{2}x_{k}^{2(\eta-1)}}\,.
\end{eqnarray}
We have calculated the mean and the variance by averaging over $10^{5}$
trajectories. Comparison of the analytic expressions (\ref{eq:avg-x}),
(\ref{eq:avg-x-2}) with numerically obtained time-dependent mean and variance is
shown in Fig.~\ref{fig:mean-var}. For numerical simulation we have chosen three
different values of the exponent $\eta$: $\eta=-\frac{1}{2}$,
$\eta=\frac{1}{2}$, and $\eta=\frac{3}{2}$ corresponding, respectively, to
subdiffusion, superdiffusion and to the localization of the particle. For each
case we have chosen a value of the parameter $\nu$ that differs from $\eta$ of
the free HDP. For all numerical simulations the initial position is $x_{0}=1$.
We see a good agreement of the numerical results with analytic expressions. As
we can see in Fig.~\ref{fig:mean-var}, the time dependence of the mean and the
variance becomes a power-law for large times and the initial position is
forgotten (for the parameters used in Fig.~\ref{fig:mean-var}(b) and (e) the
difference between the exact solution and the power-law approximation remains
constant, but the relative difference is decreasing).

\begin{figure}
\includegraphics[width=0.33\textwidth]{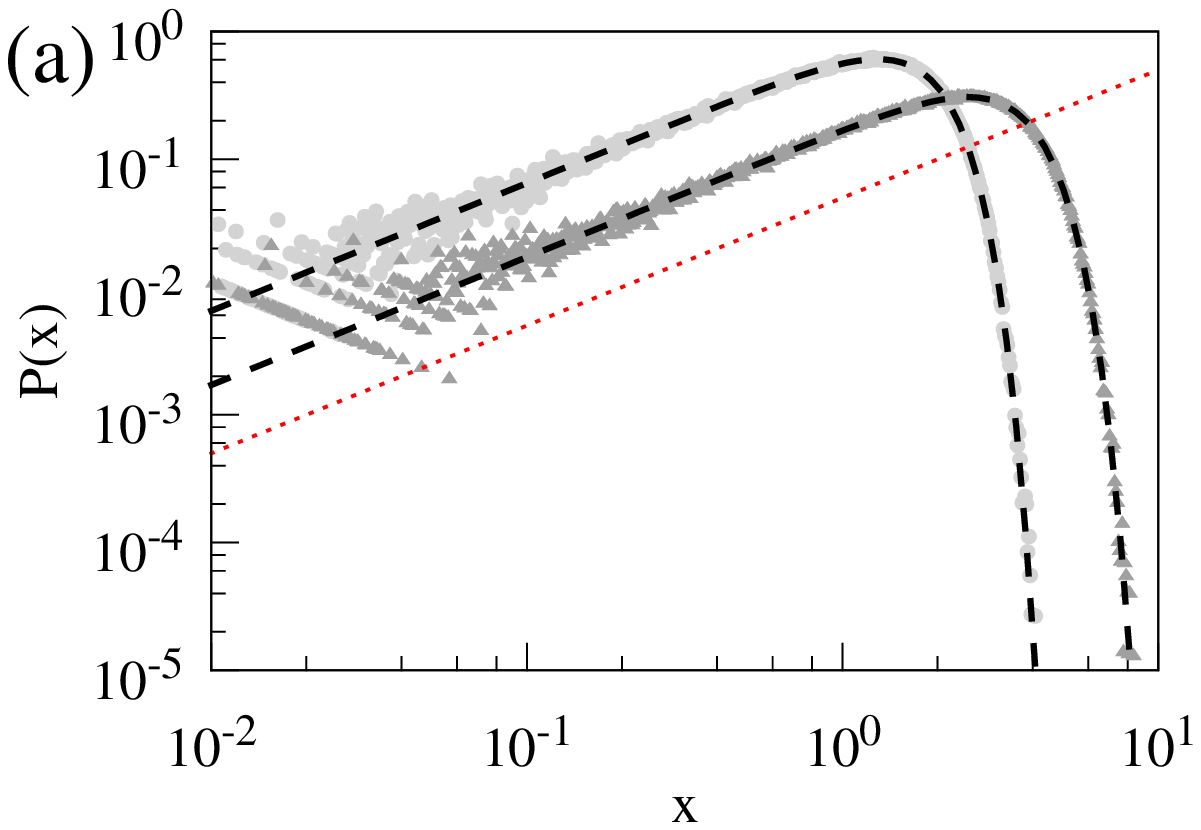}\includegraphics[width=0.33\textwidth]{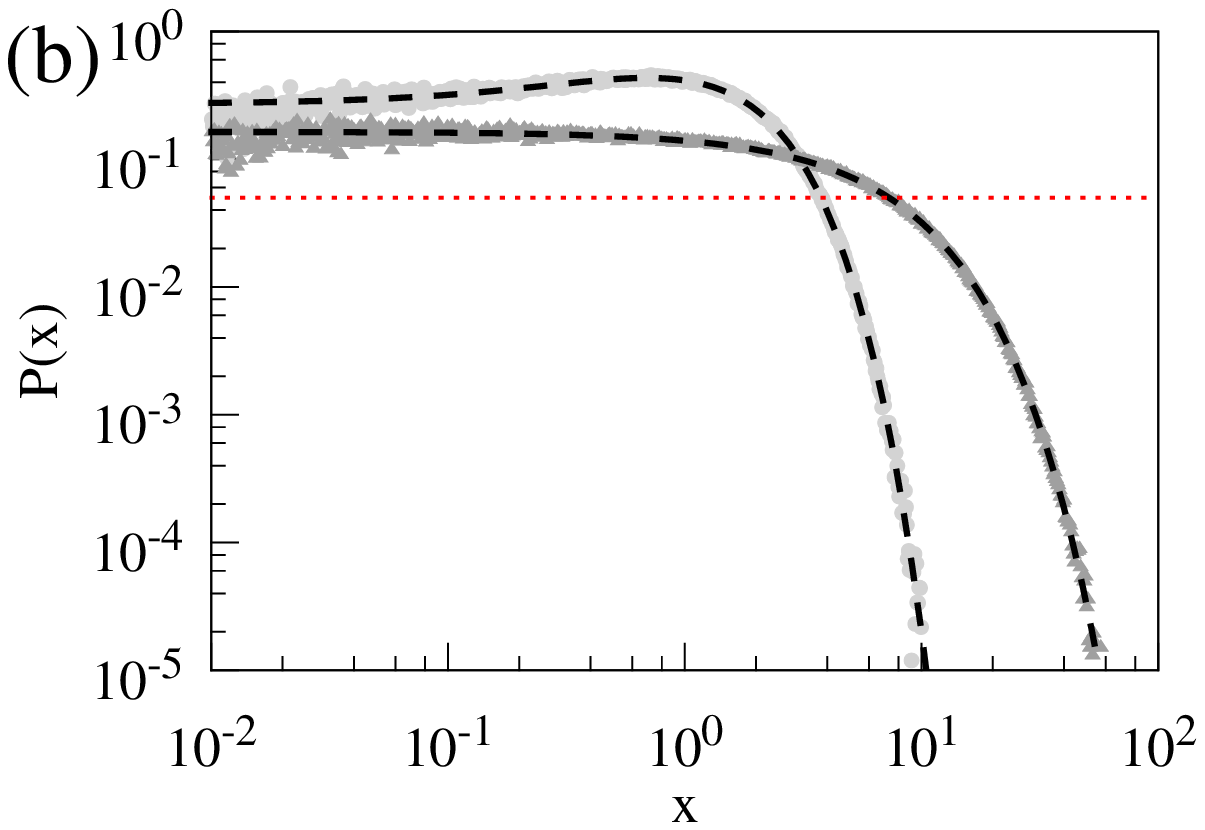}\includegraphics[width=0.33\textwidth]{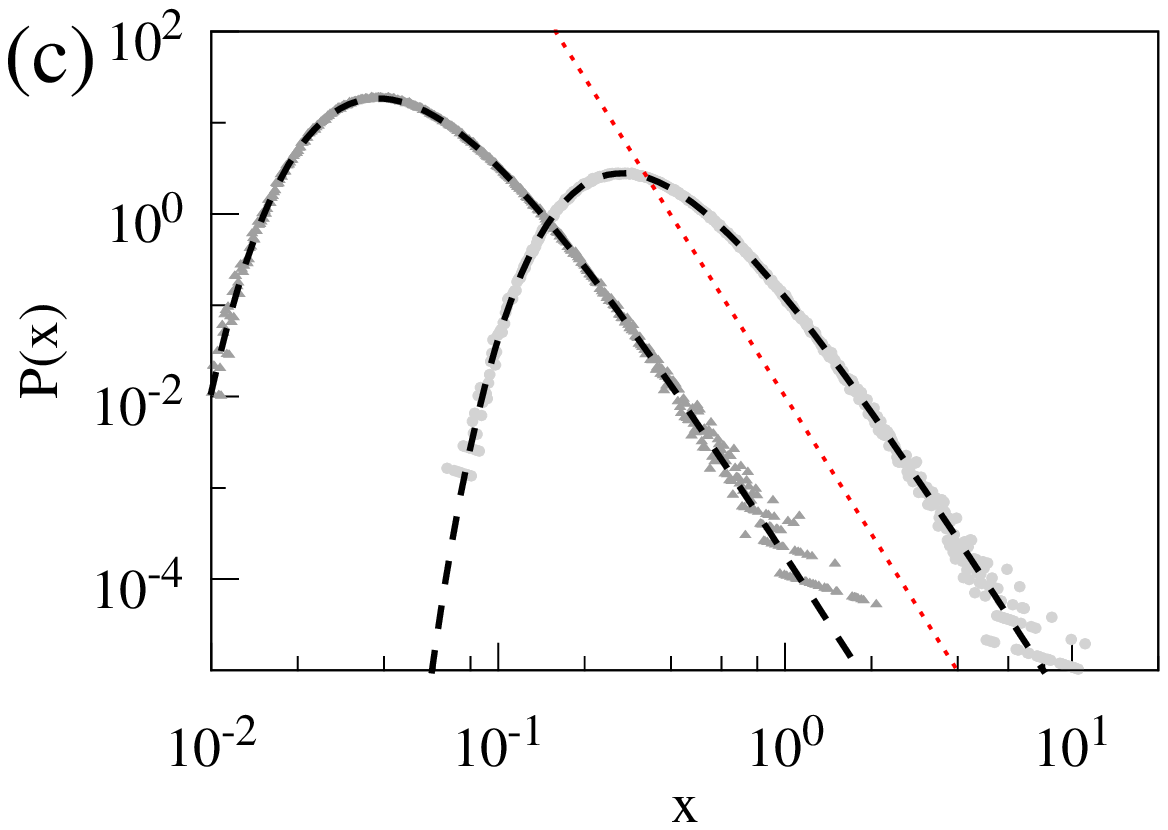}
\caption{Time-dependent PDF $P(x,t|x_{0},0)$ corresponding to times $t=1$ (light
gray) and $t=10$ (dark gray) for various values of the parameters $\eta$ and
$\nu$ when the position of the diffusing particle changes according to
Eq.~(\ref{eq:sde-1}). Dashed black lines are calculated using
Eq.~(\ref{eq:pdf-x}). The dotted line shows the slope $x^{-\nu}$. The parameters
are $\sigma=1$ and (a) $\eta=-\frac{1}{2}$, $\nu=-1$; (b) $\eta=\frac{1}{2}$,
$\nu=0$; (c) $\eta=\frac{3}{2}$, $\nu=5$. The initial position is $x_{0}=1$.}
\label{fig:pdf-1}
\end{figure}

Comparison of the analytic expression (\ref{eq:pdf-x}) for the time-dependent
PDF $P(x,t|x_{0},0)$ with the results of numerical simulation is shown in
Fig.~\ref{fig:pdf-1}. To illustrate how the PDF changes with time, the PDF is
shown at two different time moments $t=1$ and $t=10$. We see a good agreement of
the numerical results with the analytic expression for both time moments. With
increasing time the PDF shifts to the larger values of $x$ when $\eta<1$ and to
the smaller values of $x$ when $\eta>1$.

\section{External force leading to an exponential restriction of the
diffusion\label{sec:boundaries}}

Now let us consider the external deterministic force having a power-law
dependence on $x$ but with the power-law exponent different than $2\eta-1$. When
such a force is positive if the power-law exponent is smaller than $2\eta-1$ and
negative if the power-law exponent is larger than $2\eta-1$, the SDE describing
the HDP can be written as
\begin{equation}
dx=\sigma^{2}\left(\left(\eta-\frac{\nu}{2}\right)x^{2\eta-1}+
\frac{m_{1}}{2}x_{\mathrm{min}}^{m_{1}}x^{2\eta-1-m_{1}}-\frac{m_{2}}{2x_{\mathrm{max}}^{m_{2}}}
x^{2\eta-1+m_{2}}\right)dt+\sigma x^{\eta}dW_{t}\,.
\label{eq:sde-2}
\end{equation}
Here $m_{1},m_{2}>0$ and $x_{\mathrm{min}}$, $x_{\mathrm{max}}$
are the parameters of the external force. In Eq.~(\ref{eq:sde-2})
we included three terms in the drift. Each term has power-law dependence
on position $x$, but with different power-law exponents: equal to
$2\eta-1$, smaller than $2\eta-1$ and larger than $2\eta-1$. The
steady-state PDF corresponding to Eq.~(\ref{eq:sde-2}) is
\begin{equation}
P_{0}(x)\sim x^{-\nu}\exp\left\{ -\left(\frac{x_{\mathrm{min}}}{x}\right)^{m_{1}}
  -\left(\frac{x}{x_{\mathrm{max}}}\right)^{m_{2}}\right\} \,.
\end{equation}
We see that the additional terms in Eq.~(\ref{eq:sde-2}) lead to exponential
restriction of the diffusion. When $x_{\mathrm{min}}\ll x\ll x_{\mathrm{max}}$,
the steady-state PDF has the power-law form $P_{0}(x)\sim x^{-\nu}$. Confined
HDP has been investigated in Refs.~\cite{Cherstvy2014b,Cherstvy2015}. Analysis
of confined HDP can be relevant for the description of the tracer particles
moving in the confinement of cellular compartments or for the particle traced
with optical tweezers that exert a restoring force on the particle
\cite{Jeon2011}.

We can mathematically obtain one of the terms in Eq.~(\ref{eq:sde-2}) leading to
the exponential restriction of the diffusion by transforming the time in the
initial equation. Indeed, if we start with the It\^o SDE
\begin{equation}
dx=a(x)dt+b(x)dW_{t}
\end{equation}
and introduce a new stochastic variable $z(t)=g(t)x(c(t))$ then the
SDE for the stochastic variable $z$ becomes
\begin{equation}
dz=\left(\frac{dg(t)}{dt}\frac{z}{g(t)}+g(t)\frac{dc(t)}{dt}a\left(\frac{z}{g(t)}\right)\right)dt
+g(t)\sqrt{\frac{dc(t)}{dt}}b\left(\frac{z}{g(t)}\right)dW_{t}\,.
\label{eq:transform-general}
\end{equation}
In Eq.~(\ref{eq:transform-general}) a new term that is proportional
to $z$ and to the derivative of $g(t)$ appears in the drift. The
PDF of the stochastic variable $z$ is related to the PDF of $x$
via the equation
\begin{equation}
P_{z}(z,t)=\frac{1}{g(t)}P_{x}\left(\frac{z}{g(t)},c(t)\right)\,.
\label{eq:transform-pdf}
\end{equation}
Thus, to introduce a new term into Eq.~(\ref{eq:sde-1}), let us
start with a stochastic variable $y(t)$ obeying the SDE~(\ref{eq:sde-1})
and consider a new stochastic variable
\begin{equation}
x(t)=e^{\mu t}y\left(\frac{1}{\kappa}(e^{\kappa t}-1)\right)\,.
\end{equation}
Here the functions $g(t)$ and $c(t)$ are $g(t)=e^{\mu t}$ and
$c(t)=\kappa^{-1}(e^{\kappa t}-1)$. From Eq.~(\ref{eq:transform-general})
follows that the equation for the stochastic variable $x$ is
\begin{equation}
dx=\left(\mu x+\sigma^{2}\left(\eta-\frac{\nu}{2}\right)e^{\kappa t-2\mu(\eta-1)t}
x^{2\eta-1}\right)dt+\sigma e^{\frac{1}{2}\kappa t-\mu(\eta-1)t}x^{\eta}dW_{t}\,.
\end{equation}
We can obtain an equation with time-independent coefficients by requiring that
\begin{equation}
\kappa=2\mu(\eta-1)\,.
\end{equation}
Using this value for the parameter $\kappa$ the SDE for $x$ becomes
\begin{equation}
dx=\left(\mu x+\sigma^{2}\left(\eta-\frac{\nu}{2}\right)x^{2\eta-1}\right)dt
+\sigma x^{\eta}dW_{t}\,.
\end{equation}
When $\mu$ has the same sign as $\eta-1$, this SDE can be written
in the form similar to Eq.~(\ref{eq:sde-2}):
\begin{equation}
dx=\sigma^{2}\left(\eta-\frac{\nu}{2}+(\eta-1)\left(\frac{x_{\mathrm{m}}}{x}\right)^{2(\eta-1)}\right)
x^{2\eta-1}dt+\sigma x^{\eta}dW\,,\label{eq:sde-3}
\end{equation}
where the parameter $x_{\mathrm{m}}$ is defined by the equation
\begin{equation}
\mu=\sigma^{2}(\eta-1)x_{\mathrm{m}}^{2(\eta-1)}\,.
\end{equation}
Comparing Eq.~(\ref{eq:sde-3}) and Eq.~(\ref{eq:sde-2}) we see
that the time transformation considered in this Section introduces
an exponential restriction of the diffusion at small values of $x$
when $\eta>1$ and at large values of $x$ when $\eta<1$.

Using Eqs.~(\ref{eq:pdf-x}) and (\ref{eq:transform-pdf}) we obtain
the time-dependent PDF for the stochastic variable $x$ obeying SDE~(\ref{eq:sde-3}):
\begin{eqnarray}
P(x,t|x_{0},0) & = & \frac{2|\eta-1|x_{\mathrm{m}}^{2(\eta-1)}}{1-e^{-2\mu(\eta-1)t}}
x^{\frac{1-\nu-2\eta}{2}}x_{0}^{\frac{1+\nu-2\eta}{2}}e^{\frac{1+\nu-2\eta}{2}\mu t}\nonumber \\
&  & \times\exp\left(-\frac{x_{\mathrm{m}}^{2(\eta-1)}}{1-e^{-2(\eta-1)\mu t}}
\left(x^{2(1-\eta)}+x_{0}^{2(1-\eta)}e^{-2(\eta-1)\mu t}\right)\right)\nonumber \\
&  & \times I_{\frac{1+\nu-2\eta}{2(\eta-1)}}\left(\frac{x_{\mathrm{m}}^{2(\eta-1)}
x^{(1-\eta)}x_{0}^{(1-\eta)}}{\sinh\left((\eta-1)\mu t\right)}\right)
\label{eq:pdf-2}
\end{eqnarray}
The conditions of validity of this expression is the same as for Eq.~(\ref{eq:pdf-x}).
That is, the expression for the PDF given by Eq.~(\ref{eq:pdf-2})
is valid when $\nu>1$ and $\eta>1$ or $\nu<1$ and $\eta<1$. The
average of a power of $x$, calculated using Eq.~(\ref{eq:pdf-2}),
is
\begin{eqnarray}
\langle x^{a}\rangle_{x_{0}} & = & \int_{0}^{\infty}x^{a}P(x,t|x_{0},0)dy\nonumber \\
 & = & \frac{\Gamma\left(\frac{\nu-1-a}{2(\eta-1)}\right)}{\Gamma\left(\frac{\nu-1}{2(\eta-1)}\right)}
 \frac{x_{\mathrm{m}}^{a}}{(1-e^{-2(\eta-1)\mu t})^{\frac{a}{2(\eta-1)}}}\,{}_{1}F_{1}
 \left(\frac{a}{2(\eta-1)};\frac{\nu-1}{2(\eta-1)};
   -\frac{x_{\mathrm{m}}^{2(\eta-1)}x_{0}^{2(1-\eta)}}{e^{2(\eta-1)\mu t}-1}\right)\,.
 \label{eq:bound-avg-x-a}
\end{eqnarray}
This average is finite under the same conditions as Eq.~(\ref{eq:avg-x-a}).
In particular, the average of $x$ is equal to
\begin{equation}
\langle x\rangle_{x_{0}}=\frac{\Gamma\left(\frac{\nu-2}{2(\eta-1)}\right)}{\Gamma\left(\frac{\nu-1}{2(\eta-1)}\right)}
\frac{x_{\mathrm{m}}}{(1-e^{-2(\eta-1)\mu t})^{\frac{1}{2(\eta-1)}}}\,{}_{1}F_{1}
\left(\frac{1}{2(\eta-1)};\frac{\nu-1}{2(\eta-1)};
  -\frac{x_{\mathrm{m}}^{2(\eta-1)}x_{0}^{2(1-\eta)}}{e^{2(\eta-1)\mu t}-1}\right)
\label{eq:bound-avg-x}
\end{equation}
and is finite when $\nu>2$ and $\eta>1$ or $\nu<1$ and $\eta<1$.
The average of the square of $x$ is equal to
\begin{equation}
\langle x^{2}\rangle_{x_{0}}=\frac{\Gamma\left(\frac{\nu-3}{2(\eta-1)}\right)}{\Gamma\left(\frac{\nu-1}{2(\eta-1)}\right)}
\frac{x_{\mathrm{m}}^{2}}{(1-e^{-2(\eta-1)\mu t})^{\frac{1}{\eta-1}}}\,{}_{1}F_{1}
\left(\frac{1}{(\eta-1)};\frac{\nu-1}{2(\eta-1)};
  -\frac{x_{\mathrm{m}}^{2(\eta-1)}x_{0}^{2(1-\eta)}}{e^{2(\eta-1)\mu t}-1}\right)
\label{eq:bound-avg-x-2}
\end{equation}
and is finite when $\nu>3$ and $\eta>1$ or $\nu<1$ and $\eta<1$.

When $\mu$ has the same sign as $\eta-1$ and $t\rightarrow\infty$
then the PDF~(\ref{eq:pdf-2}) tends to the steady-state PDF
\begin{equation}
P_{0}(x)=\frac{2|\eta-1|x_{\mathrm{m}}^{\nu-1}}{\Gamma\left(\frac{\nu-1}{2(\eta-1)}\right)}
x^{-\nu}\exp\left(-\left(\frac{x_{\mathrm{m}}}{x}\right)^{2(\eta-1)}\right)\,.
\label{eq:pdf-2-st}
\end{equation}
The steady-state PDF (\ref{eq:pdf-2-st}) leads to the steady-state
averages of $x$ and $x^{2}$
\begin{eqnarray}
\langle x\rangle_{\mathrm{st}} & = & \frac{\Gamma\left(\frac{\nu-2}{2(\eta-1)}\right)}{\Gamma\left(\frac{\nu-1}{2(\eta-1)}\right)}
x_{\mathrm{m}}\,,\\
\langle x^{2}\rangle_{\mathrm{st}} & = & \frac{\Gamma\left(\frac{\nu-3}{2(\eta-1)}\right)}{\Gamma\left(\frac{\nu-1}{2(\eta-1)}\right)}
x_{\mathrm{m}}^{2}\,.\label{eq:x-2-st}
\end{eqnarray}

Now let us consider the time evolution of the average $\langle x^{2}\rangle_{x_{0}}$,
given by Eq.~(\ref{eq:bound-avg-x-2}). In the case when the initial
position $x_{0}$ is far from the cut-off boundary $x_{\mathrm{m}}$
(that is $x_{0}\ll x_{\mathrm{m}}$ when $\eta<1$ and $x_{0}\gg x_{\mathrm{m}}$
when $\eta>1$), the time evolution of the average $\langle x^{2}\rangle_{x_{0}}$
can be separated into three parts. First, for small times 
\[
t\ll\frac{x_{0}^{2(1-\eta)}}{2(\eta-1)^{2}\sigma^{2}}
\]
the influence of the initial position is significant and the diffusion
is approximately normal, $\langle x^{2}\rangle_{x_{0}}$ depends linearly
on time $t$. For the intermediate times 
\[
\frac{x_{0}^{2(1-\eta)}}{2(\eta-1)^{2}\sigma^{2}}\ll t\ll
\frac{1}{2(\eta-1)\mu}=\frac{x_{\mathrm{m}}^{2(1-\eta)}}{2(\eta-1)^{2}\sigma^{2}}
\]
the exponent $e^{-2(\eta-1)\mu t}$ in Eq.~(\ref{eq:bound-avg-x-2})
differs from $1$ only slightly, however the last argument of the
hypergeometric function is already small. Approximating the hypergeometric
function by $1$ and expanding the exponent $e^{-2(\eta-1)\mu t}$
into power series and keeping only the linear term we obtain that
the average $\langle x^{2}\rangle_{x_{0}}$ depends on time as a power-law,
$\langle x^{2}\rangle_{x_{0}}\sim t^{1/(1-\eta)}$. Thus for this
intermediate range of time the anomalous diffusion occurs. Finally,
for large times
\[
t\apprge\frac{1}{2(\eta-1)\mu}
\]
the cut-off position $x_{\mathrm{m}}$ starts to influence the diffusion
and $\langle x^{2}\rangle_{x_{0}}$ approaches the steady-state value
(\ref{eq:x-2-st}). We can conclude that the introduction of the boundary
via an exponential cut-off does not change the anomalous diffusion
when the starting position is far from the boundary ant the time is
not too large.

\begin{figure}
\includegraphics[width=0.33\textwidth]{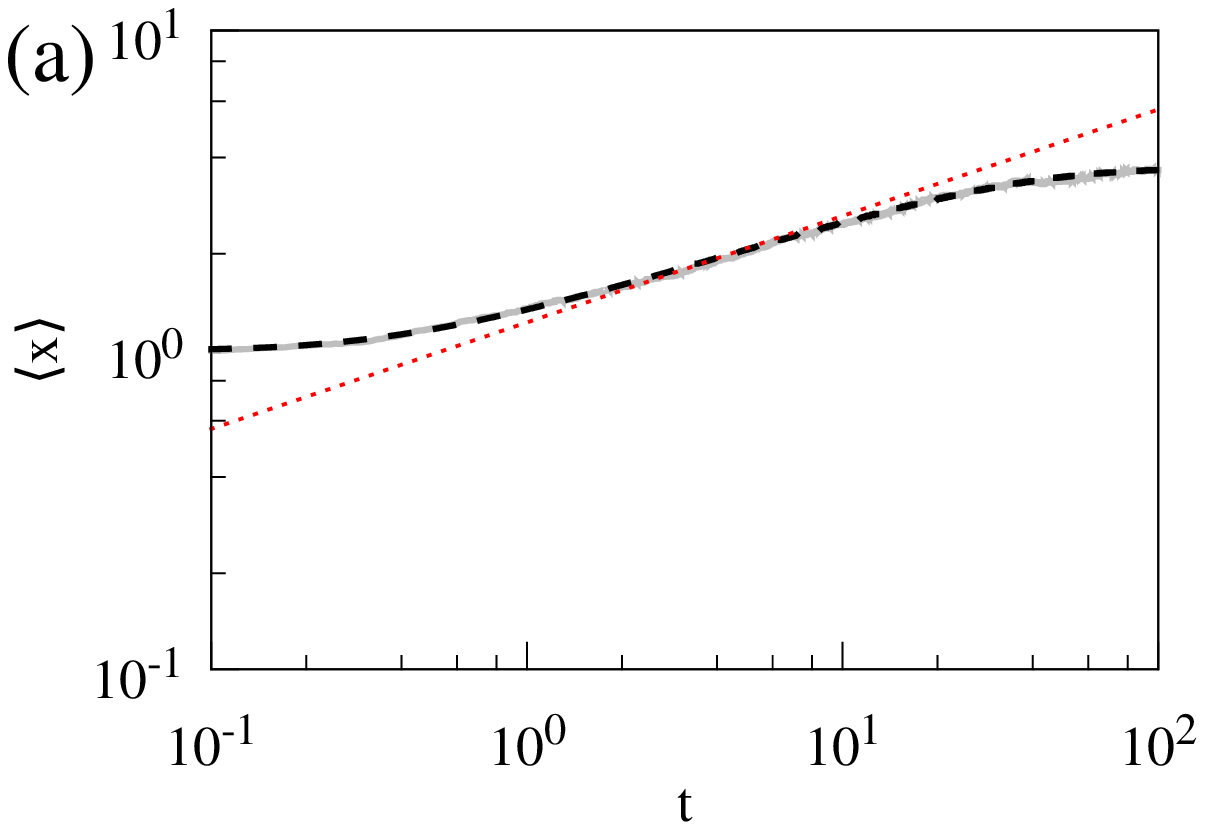}\includegraphics[width=0.33\textwidth]{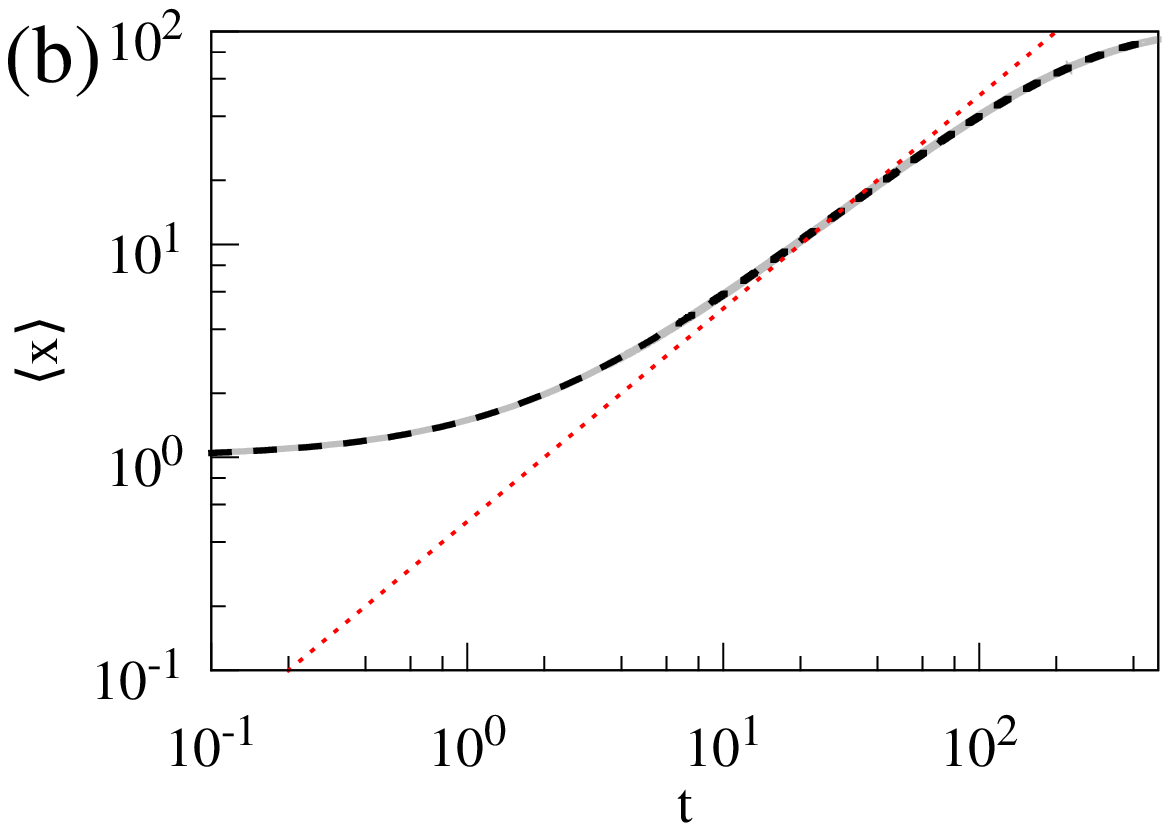}\includegraphics[width=0.33\textwidth]{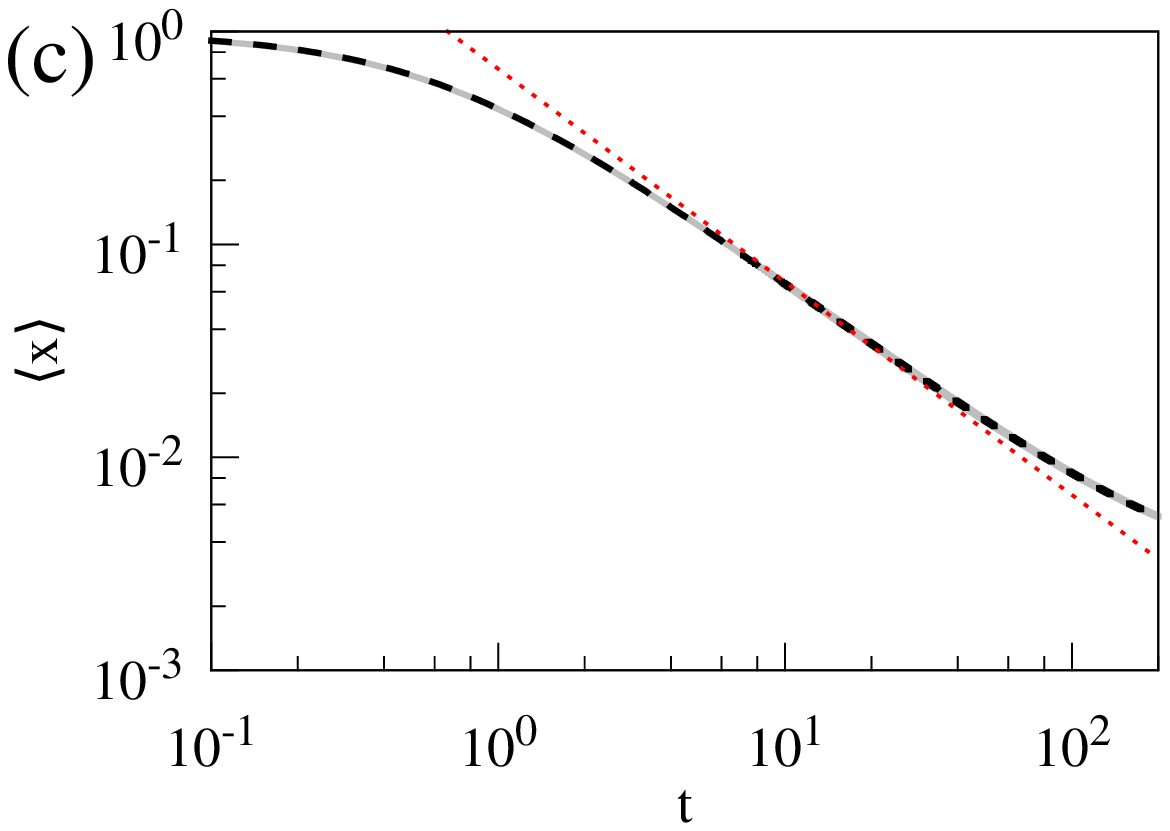}\\
\includegraphics[width=0.33\textwidth]{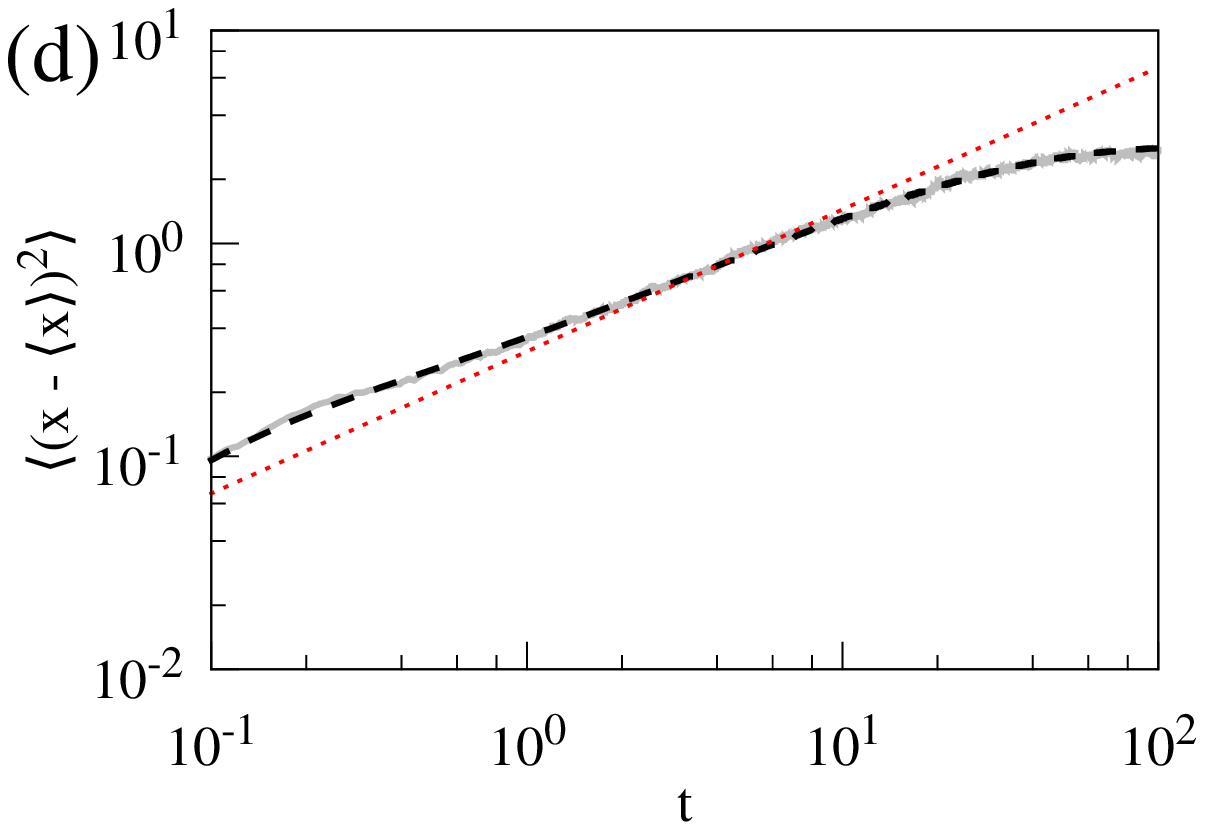}\includegraphics[width=0.33\textwidth]{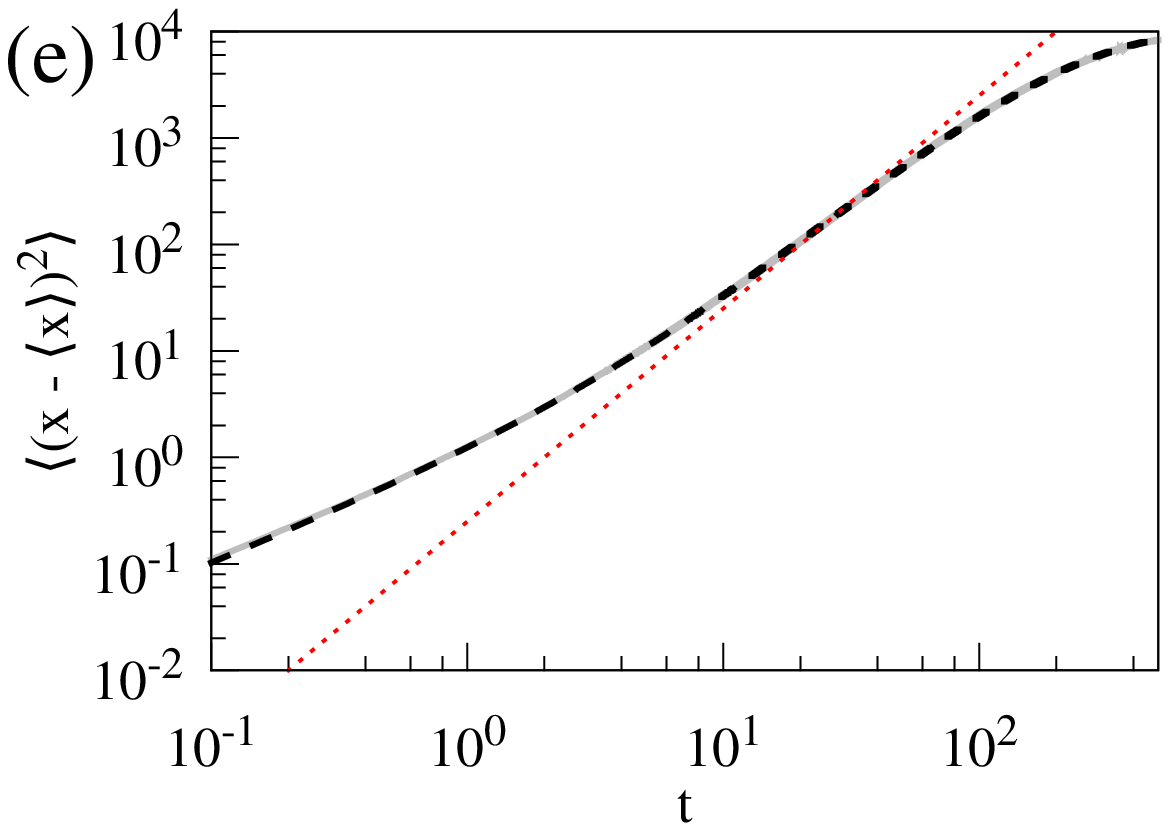}\includegraphics[width=0.33\textwidth]{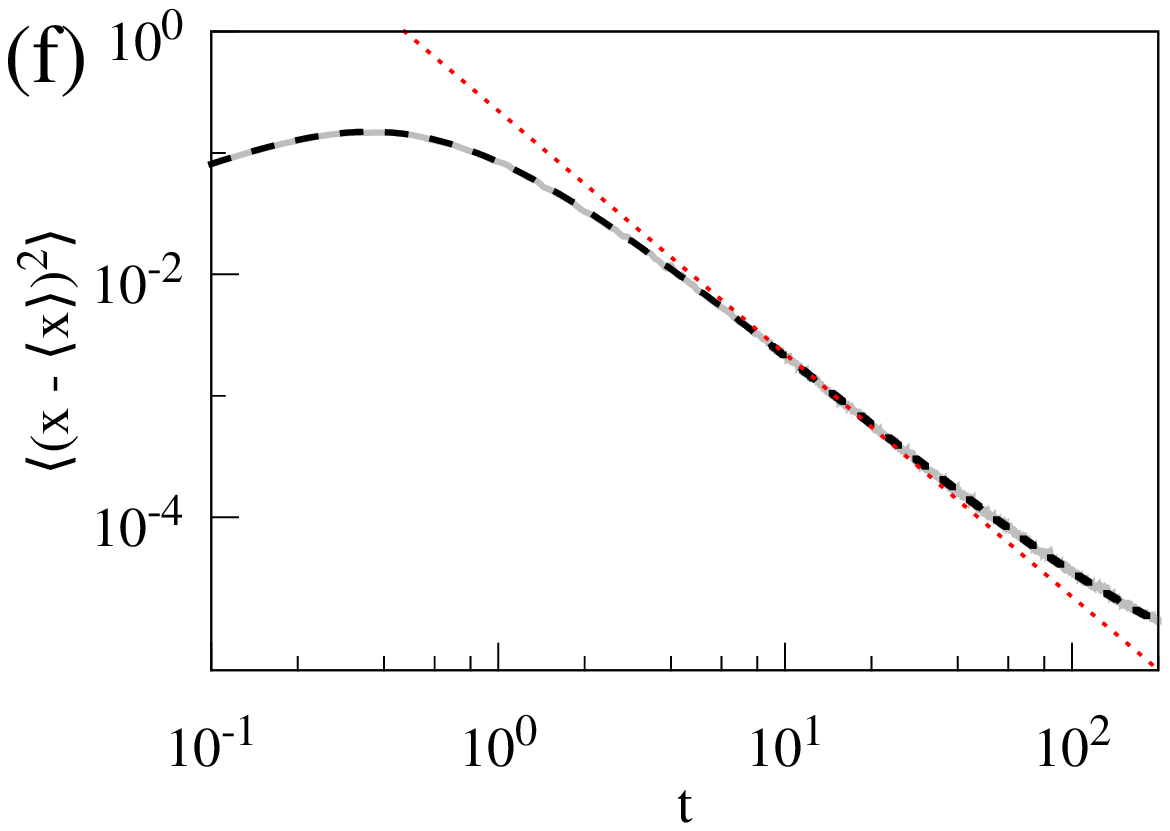}
\caption{Dependence of the mean (a,b,c) and variance (d,e,f) on time for various
values of the parameters $\eta$ and $\nu$ when the position of the diffusing
particle changes according to Eq.~(\ref{eq:sde-3}). Solid gray lines show
numerical result, dashed black lines are calculated using
Eqs.~(\ref{eq:bound-avg-x}) and (\ref{eq:bound-avg-x-2}), dotted (red) lines
show the power-law dependence on time $\sim t^{1/[2(1-\eta)]}$ for (a,b,c) and
$\sim t^{1/(1-\eta)}$ for (d,e,f). The parameters are $\sigma=1$ and
$\eta=-\frac{1}{2}$, $\nu=-1$, $x_{\mathrm{m}}=5$ for (a,d); $\eta=\frac{1}{2}$,
$\nu=0$, $x_{\mathrm{m}}=100$ for (b,c); $\eta=\frac{3}{2}$, $\nu=5$
$x_{\mathrm{m}}=0.01$ for (c,f). The initial position is $x_{0}=1$.}
\label{fig:mean-var-bound}
\end{figure}

Comparison of the analytic expressions (\ref{eq:bound-avg-x}),
(\ref{eq:bound-avg-x-2}) with numerically obtained time-dependent mean and
variance is shown in Fig.~\ref{fig:mean-var-bound}. As in Sec.~\ref{sec:drift},
for numerical solution we use the Euler-Maruyama scheme with a variable time
step, equivalent to the introduction of the operational time in addition to the
physical time $t$. When the diffusion coefficient in the operational time $\tau$
does not depend on position $x$, the numerical method is given by the equations
\begin{eqnarray}
x_{k+1} & = & x_{k}+\left[\left(\eta-\frac{\nu}{2}\right)\frac{1}{x_{k}}
  +(\eta-1)x_{\mathrm{m}}^{2(\eta-1)}x_{k}^{1-2\eta}\right]\Delta\tau
+\sqrt{\Delta\tau}\varepsilon_{k}\,,\\
t_{k+1} & = & t_{k}+\frac{\Delta\tau}{\sigma^{2}x_{k}^{2\eta}}\,.
\end{eqnarray}
When $\eta>1$, a more efficient numerical method is obtained when
the change of the variable $x$ in one step is proportional to the
value of the variable. Then the numerical method of solution is described
by the equations
\begin{eqnarray}
x_{k+1} & = & x_{k}\left[1+\Delta\tau\left(\eta-\frac{\nu}{2}\right)
  +\Delta\tau(\eta-1)\left(\frac{x_{\mathrm{m}}}{x_{k}}\right)^{2(\eta-1)}
  +\sqrt{\Delta\tau}\varepsilon_{k}\right]\,,\\
t_{k+1} & = & t_{k}+\frac{\Delta\tau}{\sigma^{2}x_{k}^{2(\eta-1)}}\,.
\end{eqnarray}
When $\eta<1$, to avoid the divergence of the diffusion and drift coefficients
at $x=0$ in the numerical simulation, we insert a reflective boundary at
$x=10^{-3}$ . This is not necessary when $\eta>1$ because the additional term in
the drift creates an exponential cut-off at small values of $x\sim
x_{\mathrm{m}}$. For numerical simulation we used the same values of the
parameters $\eta$ and $\nu$ as in Fig.~\ref{fig:mean-var}, the initial position
is $x_{0}=1$. We have chosen the parameter $x_{\mathrm{m}}$ of the external
force so that the initial position $x_{0}$ is far from the boundary
$x_{\mathrm{m}}$. In Fig.~\ref{fig:mean-var-bound} we see a good agreement of
the numerical results with analytic expressions. The numerical calculation and
the analytic expressions confirm the presence of a time interval where the mean
and the variance have a power-law dependence on time, as can be seen in
Fig.~\ref{fig:mean-var-bound}. The upper limit of this time interval is
determined by the the position $x_{\mathrm{m}}$ of the exponential cut-off.

\begin{figure}
\includegraphics[width=0.33\textwidth]{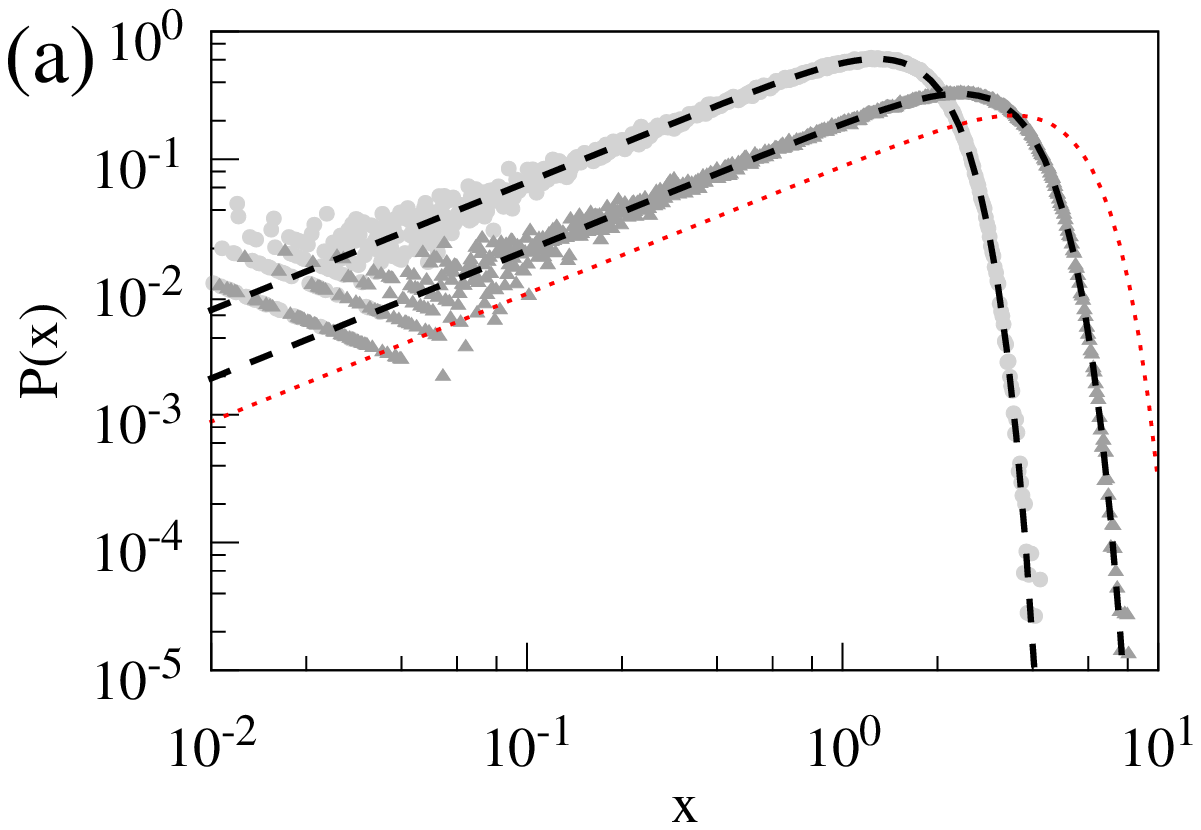}\includegraphics[width=0.33\textwidth]{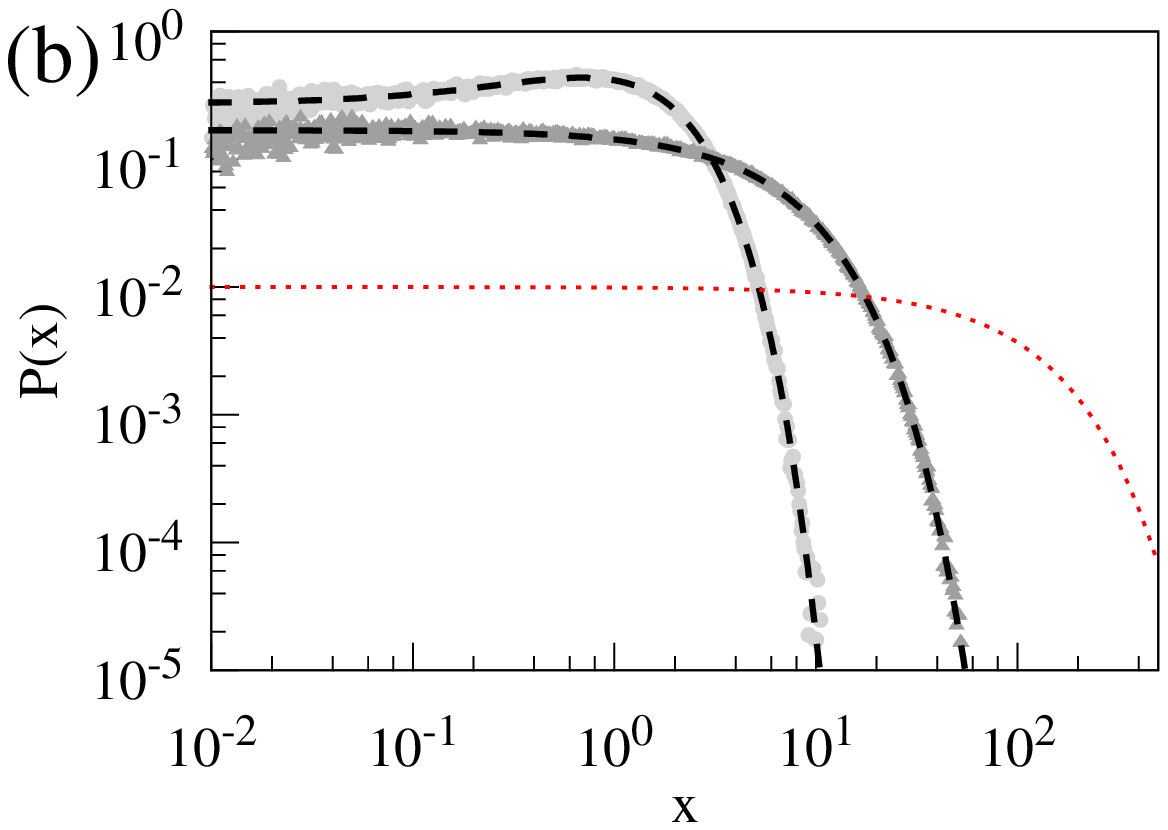}\includegraphics[width=0.33\textwidth]{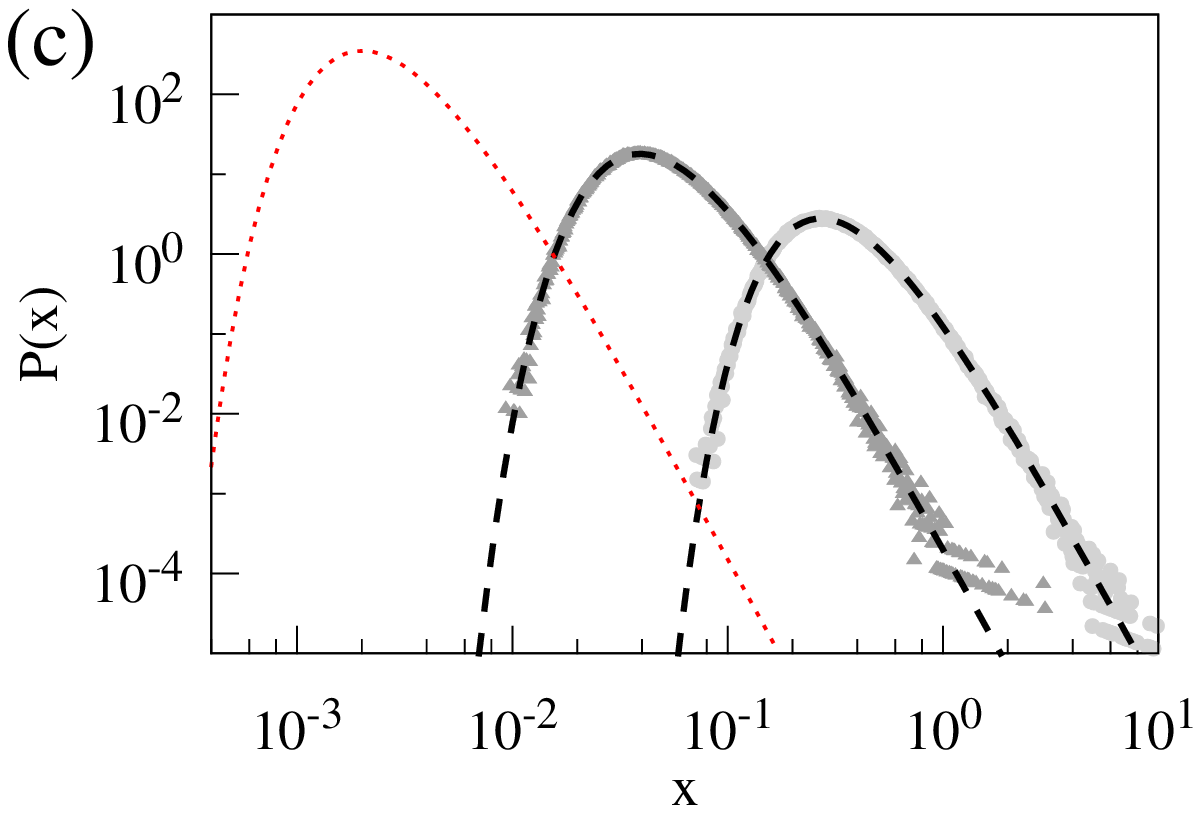}
\caption{Time-dependent PDF $P(x,t|x_{0},0)$ corresponding to times $t=1$ (light
gray) and $t=10$ (dark gray) for various values of the parameters $\eta$ and
$\nu$ when the position of the diffusing particle changes according to
Eq.~(\ref{eq:sde-3}). Dashed black lines are calculated using
Eq.~(\ref{eq:pdf-2}). The dotted line shows the steady-state PDF
(\ref{eq:pdf-2-st}). The parameters are $\sigma=1$ and (a) $\eta=-\frac{1}{2}$,
$\nu=-1$, $x_{\mathrm{m}}=5$; (b) $\eta=\frac{1}{2}$, $\nu=0$,
$x_{\mathrm{m}}=100$; (c) $\eta=\frac{3}{2}$, $\nu=5$ $x_{\mathrm{m}}=0.01$. The
initial position is $x_{0}=1$.}
\label{fig:pdf-2}
\end{figure}

Comparison of the analytic expression (\ref{eq:pdf-2}) for the time-dependent
PDF $P(x,t|x_{0},0)$ with the results of numerical simulation is shown in
Fig.~\ref{fig:pdf-2}. We see a good agreement of the numerical results with the
analytic expression. With increasing time the PDF shifts to the larger values of
$x$ when $\eta<1$ and to the smaller values of $x$ when $\eta>1$. However, in
contrast to the situation in the previous Section where the restricting force
was not present, this shift of the PDF now is limited, at large times the
time-depend PDF approaches the steady-state PDF (\ref{eq:pdf-2-st}).

\section{Conclusions\label{sec:concl}}

In summary, we have obtained analytic expressions (\ref{eq:pdf-x}) and
(\ref{eq:pdf-2}) for the transition probability of the heterogeneous diffusion
process whose diffusivity has a power-law dependence on the distance $x$. In the
description of the HDP we have included an additional deterministic force that
also has a power-law dependence on the position. The drift term having a
power-law dependence on the position can arise not only due to an external force
but can also represent a noise-induced drift \cite{Volpe2016}. Such a drift term
appears in a Langevin equation describing overdamped fluctuations of the
position of a particle in nonhomogeneous medium \cite{Sancho1982}. Stochastic
differential equations with power-law drift and diffusion terms have been used
to model random fluctuations of the atmospheric forcing on the ocean circulation
\cite{Ditlevsen1999a} and pressure time series routinely used to define the
index characterizing the North Atlantic Oscillation \cite{Lind2005}. The
Brownian motion of a colloidal particle in water subjected to the gravitational
force and with a space-dependent diffusivity due to the presence of the bottom
wall of the sample cell has been investigated in Ref.~\cite{Volpe2010}. Force
causing exponential cut-off in the PDF of the particle position can describe HDP
process in confined regions; such a description can be relevant, for example,
for the tracer particles moving in the confinement of cellular compartments
\cite{Jeon2011,Kuehn2011}.

A system obeying SDEs with power-law drift and diffusion terms can be
experimetally realized as an electrical circuit driven by a multiplicative
noise, similarly as in Ref.~\cite{Smythe1983}. The equation describing the
overdamped motion of the Brownian particle takes the form of
Eq.~(\ref{eq:sde-1}) when a temperature gradient is present in the medium and
the particle is subjected to the external potential that is proportional to the
temperature profile \cite{Kazakevicius2015}. In particular, steady state heat
transfer due to the temperature difference between the beginning and the end of
the system corresponds to $\eta=1/2$ in Eq.~(\ref{eq:sde-1})
\cite{Kazakevicius2015}. For a charged particle the external force can be
introduced by applying the electric potential difference at the ends of the
system.

When the power-law exponent in the deterministic force is equal to $2\eta-1$,
where $2\eta$ is the power-law exponent in the dependence of the diffusion
coefficient on the position, the external force does not limit the region of
diffusion. Other values of the power-law exponent in the deterministic force
can cause the exponential cut-off in the PDF of the particle positions. Such an
exponential restriction of the diffusion appears when the external force is
positive if the power-law exponent is smaller than $2\eta-1$ and negative if
the power-law exponent is larger than $2\eta-1$.  We obtained an analytic
expression (\ref{eq:pdf-2}) for the transition probability in a particular case
when the external restricting force has a linear dependence on the position.
Using analytic expression for the transition probability we calculated the time
dependence of the moments of the particle position, Eqs.~(\ref{eq:avg-x-a}) and
(\ref{eq:bound-avg-x-a}).

We found that the power-law exponent in the dependence of the MSD on time does
not depend on the external force, this force changes only the anomalous
diffusion coefficient. This conclusion is valid for sufficiently large times
satisfying the condition (\ref{eq:cond-t-large}), that is when the initial
position of the particle is forgotten and the anomalous diffusion occurs. In
addition, the external force having the power-law exponent different from
$2\eta-1$ limits the time interval where the anomalous diffusion occurs. The
conclusions remain valid also when the external force can be represented as a
sum of several terms, each term being a power-law function of position with a
different power-law exponent. Thus, our results indicate that the anomalous
diffusion caused by diffusivity being a power-law function of the position is
robust with respect to an external perturbation, the exponent $\alpha$ in
Eq.~(\ref{eq:anomalous-diff}) is determined only by the diffusion coefficient.

In addition, the results of Sec.~\ref{sec:drift} show that the character of the
anomalous diffusion does not depend on the interpretation of the Langevin
equation: the scaling exponent $\alpha$ in Eq.~(\ref{eq:anomalous-diff}) is the
same for both Stratonovich and It\^o conventions. This is because different
interpretations correspond to the different values of the parameter $\nu$ in
Eq.~(\ref{eq:sde-1}): Stratonovich convention results in $\nu=\eta$, It\^o
convention in $\nu=2\eta$, and the scaling exponent $\alpha$ does not depend on
$\nu$. The same conclusion that the exponent of the anomalous diffusion does not
depend on the prescription has been obtained in
Refs.~\cite{Fa2003,Fa2005,Heidernatsch2015} for equations describing diffusion
without the presence of an external force.

\appendix

\section{Solution of the Fokker-Planck equation for the Bessel process}

\label{sec:bessel}Let us consider the Fokker-Planck equation
\begin{equation}
\frac{\partial}{\partial t}P=\frac{\nu}{2}\frac{\partial}{\partial x}
\frac{1}{x}P+\frac{1}{2}\frac{\partial^{2}}{\partial x^{2}}P\label{eq:FP-A}
\end{equation}
and search for the time-dependent solution $P(x,t|x_{0},0)$ with
the initial condition $P(x,0|x_{0},0)=\delta(x-x_{0})$. The unnormalized
time-independent solution of Eq.~(\ref{eq:FP-A}) is $x^{-\nu}$.
The boundary condition at $x=0$ for Eq.~(\ref{eq:FP-A}) can be
expressed using the probability current \cite{Risken1989}
\begin{equation}
S(x,t)=-\frac{\nu}{2x}P(x,t)-\frac{1}{2}\frac{\partial}{\partial x}P(x,t)\,.
\label{eq:prob-current}
\end{equation}
We consider the boundary condition corresponding to the vanishing
probability current at $x=0$, $S(0,t)=0$. 

One of the possible ways to obtain the solution of Eq.~(\ref{eq:FP-A})
is to use a Laplace transform \cite{Jeanblanc2009}. Here we solve
Eq.~(\ref{eq:FP-A}) using the method of eigenfunctions. This method
has been used in Ref.~\cite{Ruseckas2010} for an equation, similar
to Eq.~(\ref{eq:FP-A}). An ansatz of the form
\begin{equation}
P(x,t)=P_{\lambda}(x)e^{-\lambda t}
\end{equation}
leads to the equation
\begin{equation}
\frac{\nu}{2}\frac{\partial}{\partial x}\frac{1}{x}P_{\lambda}
+\frac{1}{2}\frac{\partial^{2}}{\partial x^{2}}P_{\lambda}=
-\lambda P_{\lambda}\,,\label{eq:eigen}
\end{equation}
where $P_{\lambda}(x)$ are the eigenfunctions and $\lambda\geqslant0$
are the corresponding eigenvalues. The eigenfunctions obey the orthonormality
relation \cite{Risken1989}
\begin{equation}
\int_{0}^{\infty}x^{\nu}P_{\lambda}(x)P_{\lambda'}(x)\, dx=\delta(\lambda-\lambda')\,.
\label{eq:orthonorm}
\end{equation}
Expansion of the transition probability density $P(x,t|x_{0},0)$
in terms of the eigenfunctions has the form \cite{Risken1989}
\begin{equation}
P(x,t|x_{0},0)=\int_{0}^{\infty}P_{\lambda}(x)x_{0}^{\nu}P_{\lambda}(x_{0})
e^{-\lambda t}d\lambda\,.
\label{eq:expansion}
\end{equation}
For the solution of Eq.~(\ref{eq:eigen}) it is convenient to write
the eigenfunctions $P_{\lambda}(x)$ as
\begin{equation}
P_{\lambda}(x)=x^{1-\gamma}u_{\lambda}(x)\,,\label{eq:p-u}
\end{equation}
where
\begin{equation}
\gamma=\frac{1+\nu}{2}\,.
\end{equation}
Similar anzatz has been used in Refs.~\cite{Bray2000,Martin2011}.
The functions $u_{\lambda}(x)$ obey the equation
\begin{equation}
x^{2}\frac{d^{2}}{dx^{2}}u_{\lambda}+x\frac{d}{dx}u_{\lambda}
+(\rho^{2}x^{2}-\gamma^{2})u_{\lambda}=0\,,\label{eq:u}
\end{equation}
where
\begin{equation}
\rho=\sqrt{2\lambda}\,.
\end{equation}
The probability current $S_{\lambda}(x)$, Eq.~(\ref{eq:prob-current}),
rewritten in terms of functions $u_{\lambda}$, becomes
\begin{equation}
S_{\lambda}(x)=-\frac{1}{2}x^{-\gamma}\left(\gamma u_{\lambda}(x)
  +x\frac{d}{dx}u_{\lambda}(x)\right)\,.\label{eq:prob-current-2}
\end{equation}
The orthonormality of eigenfunctions (\ref{eq:orthonorm}) yields
the orthonormality for functions $u_{\lambda}(x)$
\begin{equation}
\int_{0}^{\infty}xu_{\lambda}(x)u_{\lambda'}(x)\, dx=\delta(\lambda-\lambda')\,.
\label{eq:orthonorm-2}
\end{equation}
The general solution of Eq.~(\ref{eq:u}) is
\begin{equation}
u_{\lambda}(x)=c_{1}J_{\gamma}(\rho x)+c_{2}Y_{\gamma}(\rho x)\,,\label{eq:u-2}
\end{equation}
where $J_{\gamma}(x)$ and $Y_{\gamma}(x)$ are the Bessel functions
of the first and second kind, respectively. The coefficients $c_{1}$
and $c_{2}$ needs to be determined from the boundary and normalization
conditions for the functions $u_{\lambda}(x)$. Using Eqs.~(\ref{eq:u-2})
and (\ref{eq:prob-current-2}) we get the probability current
\begin{equation}
S_{\lambda}(x)=-\frac{1}{2}\rho x^{1-\gamma}[c_{1}J_{\gamma-1}(\rho x)
+c_{2}Y_{\gamma-1}(\rho x)]\,.
\end{equation}
The requirement $S_{\lambda}(0)=0$ leads to the condition
$c_{2}=-c_{1}\tan(\pi\gamma)$. Taking into account this relation between $c_{2}$
and $c_{1}$ we obtain
\begin{equation}
u_{\lambda}(x)=c_{\lambda}J_{-\gamma}(\rho x)\,.\label{eq:u-3}
\end{equation}
In addition, the condition $S_{\lambda}(0)=0$ implies $c_{2}=0$
when $\gamma\geqslant1$. Thus, when $\gamma\geqslant1$, $\nu\geqslant1$
both coefficients $c_{1}$ and $c_{2}$ are zero and the solution
(\ref{eq:expansion}) is not valid. It is known that for a Bessel
process with such parameters a total absorption at the origin occurs
in a finite time \cite{Karlin1981}.

Orthonormality condition (\ref{eq:orthonorm-2}) leads to the equation
\begin{equation}
c_{\lambda}c_{\lambda'}\int_{0}^{\infty}xJ_{-\gamma}(\rho x)
J_{-\gamma}(\rho'x)\, dx=\delta(\lambda-\lambda')\,.
\end{equation}
Since for the Bessel functions the equality
\begin{equation}
\rho\int_{0}^{\infty}xJ_{\gamma}(\rho x)J_{\gamma}(\rho'x)xdx=\delta(\rho-\rho')
\end{equation}
is valid, we obtain $c_{\lambda}=1$. Using Eqs.~(\ref{eq:expansion}),
(\ref{eq:p-u}) and (\ref{eq:u-3}) the solution of the Fokker-Planck
equation can be expressed as
\begin{equation}
P(x,t|x_{0},0)=x^{1-\gamma}x_{0}^{\gamma}\int_{0}^{\infty}
J_{-\gamma}(\rho x)J_{-\gamma}(\rho x_{0})e^{-\frac{1}{2}\rho^{2}t}\rho d\rho\,.
\end{equation}
Integration yields
\begin{equation}
P(x,t|x_{0},0)=\frac{x^{1-\gamma}x_{0}^{\gamma}}{t}
\exp\left(-\frac{x^{2}+x_{0}^{2}}{2t}\right)I_{-\gamma}\left(\frac{xx_{0}}{t}\right)\,.
\end{equation}
Here $I_{\alpha}(x)$ is the modified Bessel function of the first
kind.

\end{document}